\documentclass[journal,10pt]{IEEEtran}

\ifCLASSINFOpdf

\else

\fi

\usepackage{CJK}
\usepackage{algorithm}
\usepackage{pifont}

\usepackage{algorithmic}
\usepackage{amsmath}
\usepackage{amssymb}
\usepackage{psfrag}
\usepackage{graphicx}
\usepackage{epstopdf}
\usepackage{multirow}
\usepackage{stfloats}
\usepackage{cite}
\usepackage{subcaption}
\usepackage{color}
\usepackage[normalem]{ulem}
\usepackage{arydshln}
\usepackage{enumerate}
\usepackage{bbm}\usepackage{verbatim}

\newtheorem{theorem}{\bf{Theorem}}[section]

\newcommand{\bm}[1]{\mbox{\boldmath{$#1$}}}

\begin{document}

\title{Learning-Based Intermittent CSI Estimation with Adaptive Intervals in Integrated Sensing and Communication Systems}

\author{Jie Chen, \IEEEmembership{Member, IEEE}, and Xianbin Wang, \IEEEmembership{Fellow, IEEE}
\thanks{Manuscript submitted to IEEE Journal of Selected Topics in Signal Processing on
 20 June 2023; revised on 26 January 2024, 23 May 2024, and 11 September 2024;  accepted on 19 September 2024.
This work was supported in part by the Natural Sciences and Engineering Research Council of Canada (NSERC) Discovery Program under Grant RGPIN-2024-05720  and in part by the Canada Research Chair Program.   (Corresponding author: Xianbin Wang.)

 J. Chen and X. Wang are with the  Department of Electrical and Computer Engineering, Western University, London, ON N6A 5B9, Canada (e-mails: chenjie.ay@gmail.com, xianbin.wang@uwo.ca).
}

}
 \maketitle

\begin{abstract}
Due to the distinct objectives and multipath utilization mechanisms between the communication and radar modules, the system design of integrated sensing and communication (ISAC) necessitates two types of channel state information (CSI), i.e., communication CSI representing the whole channel gain and phase shifts, and radar CSI exclusively focused on target mobility and position information.
However, current ISAC systems apply an identical mechanism to estimate both types of CSI at the same predetermined estimation interval based on the worst case of dynamic environments, leading to significant overhead and compromised performances.
Therefore, this paper proposes an intermittent communication and radar CSI estimation scheme with adaptive intervals for individual users/targets, where both types of CSI can be predicted using channel temporal correlations for cost reduction or re-estimated via signal transceiving for improved estimation accuracy.
Specifically, we jointly optimize the binary CSI re-estimation/prediction decisions and transmit beamforming matrices for individual users/targets to maximize communication transmission rates and minimize radar tracking errors and costs in a multiple-input single-output (MISO) ISAC system.
Unfortunately, this problem has causality issues because it requires comparing system performances under re-estimated CSI and predicted CSI during the optimization.
However, the re-estimated CSI can only be obtained after completing the optimization.
Additionally, the binary decision makes the joint design a mixed integer nonlinear programming (MINLP) problem, resulting in high complexity when using conventional optimization algorithms.
Therefore, we propose a deep reinforcement online learning (DROL) framework that first implements an online deep neural network (DNN) to learn the binary CSI updating policy from the experiences.
Given the learned policy, we propose an efficient algorithm to solve the remaining beamforming design problem.
Finally, simulation results validate the effectiveness of the proposed algorithm.
\end{abstract}
\begin{IEEEkeywords}
Integrated sensing and communication (ISAC),   channel estimation, deep reinforcement learning, neural network
\end{IEEEkeywords}

\IEEEpeerreviewmaketitle

\section{Introduction}
With the emergence of many new applications from vertical industries, future beyond 5G (B5G) and sixth-generation (6G) wireless networks are expected to provide accurate sensing and highly dependable communication services concurrently.
Integrated sensing and communication (ISAC), also called joint radar-communication (JRC) and dual-function radar-communication (DFRC), has been envisioned as the key enabler of B5G/6G to concurrently address the stringent sensing and communication demands \cite{liu2020joint,liu2022survey,zhang2021enabling,liu2022integrated}.
With integrated sensing/tracking/communication functions, shared signal waveforms, and unified physical platforms, ISAC can significantly improve overall hardware, energy, and spectrum efficiencies.
Hence, ISAC has garnered significant research interest from industry and academia worldwide in recent years.

One of the fundamental challenges in ISAC is to explore spectral sharing schemes that can achieve desirable trade-offs between communication and sensing performances~\cite{luong2021radio}.
While opportunistic spectrum sharing can provide idle frequency bands to each other \cite{zhang2022time}, these designs cannot concurrently achieve sensing and communication demands.  Hence, most research efforts in \cite{liu2020joint1,rihan2018optimum,lyu2022joint,yu2022precoding,bazzi2023outage,chen2022generalized,qian2018joint,hua2022mimo,mu2022noma,liu2018toward,li2024maximizing,hu2022low} were devoted to studying underlay spectrum sharing schemes by exploiting the spatial array gain from multiple antennas.
Specifically, the beamforming matrices were optimized in~\cite{liu2020joint1,rihan2018optimum,lyu2022joint,yu2022precoding,bazzi2023outage,chen2022generalized,qian2018joint,hua2022mimo,mu2022noma,liu2018toward}
to achieve various communication-radar performance trade-offs for multiple-input multiple-output (MIMO) ISAC systems, where communication performance metrics include transmission rate \cite{liu2020joint1,rihan2018optimum,lyu2022joint}, multi-user interference (MUI) \cite{yu2022precoding}, and outage probability~\cite{bazzi2023outage}, while radar performance metrics include sensing signal-clutter-noise ratio~\cite{chen2022generalized,qian2018joint}, Cram\'{e}r-Rao bound (CRB)~\cite{hua2022mimo}, and predefined radar beam pattern accuracy in~\cite{mu2022noma,liu2018toward}.
Moreover, the value-of-service (VoS) metric was developed in \cite{li2024maximizing} to guide fairness resource allocation for concurrent heterogeneous service provisioning in a collaborative system.
Then, building upon the above systems with orthogonal frequency division multiplexing (OFDM), the transmit waveforms were investigated in~\cite{hu2022low} to minimize the weighted communication MUI and predefined radar beam pattern error under low-peak-to-average power ratio constraints.

While the above papers  \cite{liu2020joint1,rihan2018optimum,lyu2022joint,yu2022precoding,bazzi2023outage,chen2022generalized,qian2018joint,hua2022mimo,mu2022noma,liu2018toward,li2024maximizing,hu2022low} focus on resource sharing schemes to achieve communication-sensing performance trade-offs in ISAC systems, there are a growing number of studies  \cite{liu2020radar,yuan2020bayesian,liu2022learning,wu2022resource} that devote significant efforts to exploring resource sharing mechanisms in ISAC-based tracking scenarios.
Specifically, the extended Kalman filtering (EKF) approach and message passing algorithm were proposed in \cite{liu2020radar} and \cite{yuan2020bayesian}, respectively, to
track vehicle mobility information, thus achieving predictive beamforming to improve next-frame tracking and communication performances further. Then,  a deep learning approach was proposed in \cite{liu2022learning} to learn the features of historical channels and predict next-frame beamforming matrices to maximize transmission rate under radar CRB constraints.
For a distributed heterogeneous system, the transmit power, dwell time, and bandwidth were jointly optimized in \cite{wu2022resource} to improve the tracking performance of posterior CRB (PCRB) under communication rate constraints.

Most designs in \cite{liu2020joint1,rihan2018optimum,lyu2022joint,yu2022precoding,bazzi2023outage,chen2022generalized,qian2018joint,hua2022mimo,mu2022noma,liu2018toward,li2024maximizing,hu2022low,liu2020radar,yuan2020bayesian,liu2022learning,wu2022resource}
necessitate two categories of channel state information (CSI), i.e., communication CSI representing whole channel response with channel gain and phase shift, and radar CSI revealing target angle, distance, and velocity information.
However, existing ISAC systems utilize the identical mechanism for estimating communication and radar CSI, i.e., they transmit dedicated pilots at fixed time intervals to estimate both types of CSI.
Such situation-agnostic CSI estimation schemes lead to significant training overhead and dramatically degrade system performance because they are designed based on the worst-case operation scenarios for all users and are not tailored to individual time-varying channel characteristics (e.g., changing speed and magnitude).
In fact, the commutation and radar CSI in multiple successive frames are temporally correlated due to continuous user/target moving and environment varying. By leveraging channel temporal correlations, there is no need to re-estimate CSI in each frame and the estimation interval of each user/target could be extended based on individual performance demands.

Investigating channel temporal correlation to customize CSI estimation intervals for training overhead reduction has been studied in traditional communication systems in \cite{deng2019intermittent,chopra2016throughput,lim2020efficient} and radar tracking systems in \cite{yan2015simultaneous,sun2021resource,sun2022resource}, respectively, which was further extended to ISAC systems in our previous work \cite{chen2023impact}.
Specifically, in \cite{chen2023impact}, the communication and radar CSI is only estimated in the first frame by sending training sequences, then predicted in the remaining frames by exploiting channel temporal correlations.
By deriving the impacts of channel aging time on radar CRB and communication achievable rate, the estimation intervals and other radio resources were dynamically optimized to maximize the average achievable rate subject to individual target tracking and information transmission constraints.
However, although resource allocation was applied in \cite{chen2023impact} to lengthen the estimation intervals for overhead reduction, the designed intervals are uniform for all users/targets, which actually should be customized individually for each user/target due to the different time-varying channel characteristics.
Moreover, the communication performances in \cite{deng2019intermittent,chopra2016throughput,lim2020efficient,chen2023impact} are evaluated statistically by using maximum ratio transmission (MRT) or zero-forcing (ZF)  without precise beamforming configuration according to system performance demands, thus causing system performance degradation.

To reduce training costs and improve system performance, this paper investigates the challenging problem of ISAC systems: when the communication/radar CSI of each user/target should be re-estimated according to individual channel time-varying characteristics and system performance demands.
Correspondingly, we develop the following intermittent CSI estimation scheme with adaptive intervals for a multi-user/target multiple-input single-output (MISO) ISAC system.
Specifically, binary CSI updating decisions for each user/target will be made in each frame by the base station (BS), e.g., it determines whether to re-estimate the CSI by exploiting transmitted signals with training costs or to predict it by leveraging channel temporal correlation without incurring training costs.
Next, the BS optimizes beamforming matrices for downlink ISAC signals and then transmits them to multiple users for information transmission, and simultaneously receives the echo signals reflected by targets for re-estimating the radar CSI of selected targets. However, the optimization of binary CSI updating decisions suffers from causality issues.  This is because it requires both re-estimated and predicted CSI for performance comparison, while the re-estimated CSI can only be obtained after the decision optimization has been completed. Furthermore, the binary decision optimization makes the joint design belong to a mixed integer nonlinear programming (MINLP) problem \cite{chen2023WirelessService,jia2023new,chen2020joint}, which incurs extremely high complexity. Then, we propose a deep reinforcement online learning (DROL) framework to address these issues efficiently.  To highlight the main contributions, we summarize the paper as follows:
\begin{figure}[t]
\center
\includegraphics[width=0.45\textwidth]{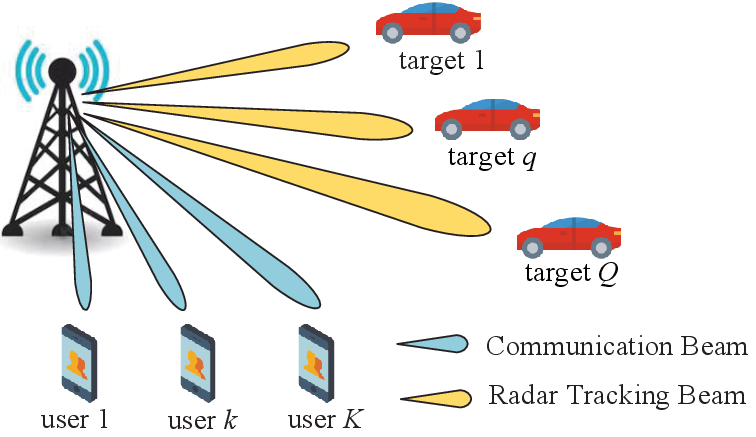}\vspace{-0.1cm}
\caption{The MISO-based ISAC system including one full-duplex BS, $K$ communication users, and $Q$  point targets. } \label{fig1}\vspace{-0.1cm}
\end{figure}

\begin{itemize}
\item
We propose an intermittent communication and radar CSI estimation scheme with adaptive intervals for individual users/targets, where both types of CSI can either be predicted using channel temporal correlations for cost efficiency or re-estimated via radio signals for improved estimation accuracy.

\item  We jointly optimize the binary CSI estimation/prediction decisions and transmit beamforming matrices for individual
users/targets to maximize the system utility, which involves the weighted achievable information transmission rate and radar performance (weighted sum of PCRB and tracking cost).

\item To avoid the causality and complexity issues mentioned above, we propose a DROL framework.
Specifically, an online deep neural network (DNN) is implemented to learn the binary CSI updating decisions from experiences in a reinforcement mechanism, thus without requiring re-estimated CSI for comparisons.
Then, with the learned CSI updating decisions, we propose an algorithm based on fractional programming (FP) \cite{guo2020weighted} and successive convex approximation (SCA) \cite{chen2019resource} to solve the beamforming problem efficiently.
\item Simulation results validate the effectiveness of the proposed scheme and show that the communication CSI with higher temporal correlation and the radar CSI with smaller state evolution noise covariance require lower estimation frequencies.
\end{itemize}

Organizations:
Section~II introduces the system model.
Section~III derives communication and radar tracking performances.
Section~IV formulates the joint CSI updating decision and beamforming optimization problem and introduces the DROL framework.
Section~V presents the details of the DROL and Section~VI shows the algorithm to solve the subproblem in the DROL.
Finally, Section~VII presents simulation results and Section~VIII concludes the paper.

Notations: ${\rm Re}\left(\cdot\right)$ and  ${\rm Im}\left(\cdot\right)$  represent the operations of taking the real   and   imaginary parts, respectively;
${\mathbb{E}}\left(\cdot\right)$ denotes the statistical expectation; ${\cal CN}\left(\mu,\sigma\right)$ denotes the circularly symmetric complex Gaussian distribution with mean $\mu$ and covariance $\sigma$; ${\bf I}_L$ denotes $L\times L$ identity matrix; $\left\lfloor \cdot \right\rfloor $  represents rounding down operation; and ${\mathbb I}_A(x)$ denotes the indicator function, i.e.,  ${\mathbb I}_A(x)=1 $ if $x\in A$, otherwise ${\mathbb I}_A(x)=0 $.

\begin{figure*}[t]
\centering
{\center\includegraphics[width= 0.98\textwidth]{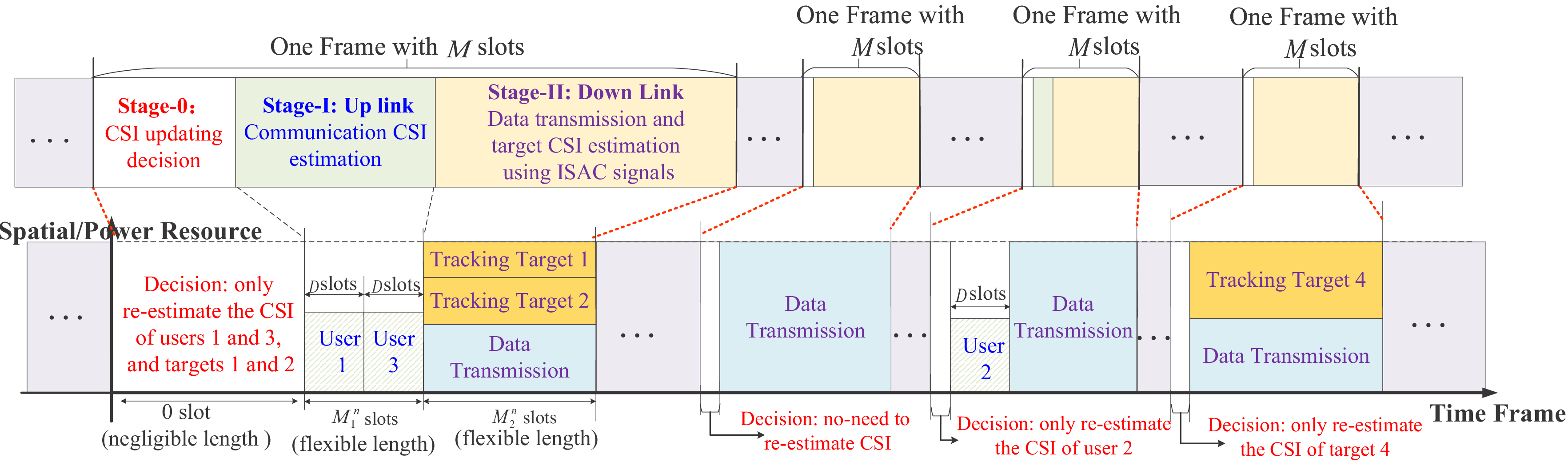} } \vspace{-0.1cm}
\caption{An illustration of intermittent CSI estimation frame structure: each frame is divided into Stage-0 (negligible length) and  Stage-I/Stage-II (flexible length). The communication CSI is intermittently estimated in Stage-I by transmitting uplink pilot sequences from users, while the radar CSI is intermittently estimated in Stage-II through the echoes of downlink beamformed ISAC signals reflected from targets.} \label{fig2}\vspace{-0.2cm}
\end{figure*}
\section{System Model}

This paper considers the MISO-based ISAC system shown in Fig.~\ref{fig1}, which consists of one full-duplex BS equipped with $L_{\rm T}$ transmit and $L_{\rm R}$ receive antennas, $K$ single-antenna communication users, and $Q$ radar tracking point targets.
To reduce training costs from communication and radar CSI estimation, we propose a unified intermittent estimation scheme for $N$ time frames, as illustrated in Fig.~\ref{fig2}.
Each frame comprising $M$ symbol slots is dynamically divided into three stages: Stage-0 for CSI updating decision-making, Stage-I for uplink communication CSI estimation, and Stage-II for downlink data transmission and target tracking (radar CSI estimation), i.e.,
\begin{itemize}
\item In Stage-0,  the BS makes binary CSI updating decisions for each user/target: either re-estimate the CSI using radio signals with training costs, or predict it by leveraging channel temporal correlation without incurring costs.
\item In Stage-I, the selected communication users transmit training sequences to the BS for communication CSI re-estimation. The training sequences for different users are transmitted over different time durations, each including $D$ symbol slots.
\item In Stage-II, the BS leverages the updated communication CSI, the predicted radar CSI, and the radar CSI updating decisions to perform beamforming design for downlink ISAC signals. Then, the BS transmits the beamformed ISAC signals to all users for individual information transmission and concurrently receives echo signals from targets to re-estimate the radar CSI of selected targets.
\end{itemize}

Note that re-estimating communication CSI incurs time consumption for pilot transmission and re-estimating radar CSI causes energy and computation costs due to successive interference cancellation (SIC) \cite{baghani2019dynamic,mei2022multi,chen2019exploiting} in the full-duplex module.
On the other hand, predicting CSI does not cause additional training costs due to without involving pilot transmission or SIC.
Moreover, the binary decisions in Stage-0 are dynamically designed according to the channel conditions and performance demands in each frame, thus the
number of users/targets whose CSI needs to be re-estimated is flexible\footnote{Note that the maximum number of communication CSI re-estimation users in one frame must not exceed $M/D$ due to the orthogonal timing allocation of training sequences in Stage-I. Also, the maximum number of radar CSI re-estimation targets in one frame is related to the number of transceiver antennas, due to the spatial multiplexing in Stage-II.}.
Hence, in specific frames, there may be no users/targets performing CSI estimation, which implies the benefits of exploiting temporal correlations for cost reduction.
Besides, we assume that the time duration in Stage-0 is negligible compared to the duration of the whole frame.

In the following, we show temporally correlated communication and radar CSI models and present the signal model for downlink data transmission and round-trip target tracking.

\subsection{Communication CSI Model}
We assume frequency-flat communication channels in the studied system~\cite{gaudio2019performance}\footnote{The proposed scheme can be extended to frequency-selective channels by adjusting the channel estimation interval for each user on each subcarrier.
However, this modification introduces complexity to both system model and algorithm design. Thus, we suggest investigating intermittent CSI estimation in frequency-selective channels as a promising topic for future research.}.
Consequently, we know that the channel responses across all subcarriers are identical.
Then, we denote the channel response on each subcarrier between the BS and the $k$-th user in the $n$-th frame by ${\bf g}_k^n$.
From \cite{chen2023impact}, the temporally correlated communication CSI model is given  by
\begin{align}
{\bf{g}}_k^n = {\rho _k}{\bf{g}}_k^{n - 1} + \sqrt {1 - \rho _k^{2}} {\bm{ \varepsilon }}_k^n\in{\mathbb C}^{L_{\rm T}\times1},\label{eqtcorrC}
\end{align}
where $ {\bm{ \varepsilon }}_k^n \sim  {\cal CN}(0,\bar \beta_k{\bf I}_{L_{\rm T}})$ represents the uncorrelated evolution noise and we assume $g_k^0 \sim  {\cal CN}(0,\bar \beta_k{\bf I}_{L_{\rm T}})$. Here,  $\bar \beta_k$ represents the large-scale fading effect. Moreover, $0\le{\rho _k}\le1$ is the prior known channel temporal correlation coefficient, which is related to its Doppler phase shift of user $k$ in Jakes' model  \cite{baddour2005autoregressive}.

\subsection{Radar CSI Model}
\begin{figure}[t]
\centering
{\center\includegraphics[width= 0.35\textwidth]{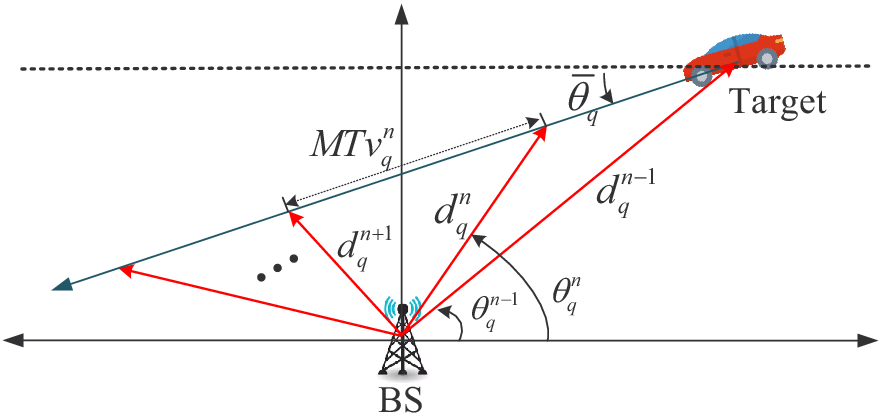} }\vspace{-0.2cm}
\caption{Radar target state evolution model.} \label{fig3} \vspace{-0.3cm}
\end{figure}

Let ${\bf{x}}_q^n = \left[ {\theta_q^n,d_q^n,v_q^n} \right]^T$ be the radar CSI of target $q$ in the $n$-th frame, where $ \theta_q^n$, $d_q^n$, and $v_q^n $ are the corresponding position angle, distance, and velocity, respectively, as shown in Fig.~\ref{fig3}. From the target state evolution model \cite{chen2023impact},  the temporally correlated radar CSI model can be given by
 \begin{align}
{\bf{x}}_q^n = \Gamma\left( {{\bf{x}}_q^{n - 1}} \right)+ {\bm \epsilon}_q^{n}\in{\mathbb C}^{3\times1},\label{eqtcorrR}
 \end{align}
where ${\bm \epsilon}_q^{n}  = \left[ {\epsilon _{qn}^\theta,\epsilon _{qn }^d,\epsilon _{qn}^v} \right]^{T}$  is the uncorrelated state evolution noise assuming ${\bm \epsilon}_q^{n } \sim {\cal C}{\cal N}\left( {0,{\bm\Sigma} _q^\epsilon } \right)$ with constant  matrix ${\bm\Sigma} _q^\epsilon  = {\rm{diag}}\left( {\epsilon_{q}^\theta, \epsilon _{q}^d, \epsilon_{q}^v} \right)$. Here, function $\Gamma(\cdot)$ is defined by
\begin{align}
{\left\{ \begin{array}{l}
\theta _q^n = \theta _q^{n - 1} + \frac{{v_q^{n - 1}MT\sin (\bar \theta _q^{n - 1})}}{{d_q^{n - 1}}}  ,\\
d_q^n = d_q^{n - 1} - v_q^{n - 1}MT\cos (\bar \theta _q^{n - 1})  ,\\
v_q^n = v_q^{n - 1}  ,
\end{array} \right.}
\label{eqARCI1}
\end{align}
where ${\bar \theta_{q}^{n-1}}={ \theta_{q}^{n-1}}-{\bar \theta_{q} }$. Here, ${\bar \theta_{q} }$ is the velocity angle with respect to
the negative horizontal direction of the BS,  and $T$ is the time duration of each symbol slot.

\subsection{ISAC Signal Model}

OFDM waveforms are widely used due to their inherent resistance to multipath fading. Besides, they can enable adaptive modulation across subcarriers while also providing significant flexibility in system design and resource management \cite{barneto2019full}.
Therefore, this paper utilizes OFDM for the ISAC design.
Specifically, the OFDM-based ISAC signals in Stage-II of the $n$-th frame  transmitted by the BS for downlink data transmission and round-trip target tracking (radar CSI estimation) is 
\begin{align}{{\bf{s}}_n}\left( t \right) =\sum\nolimits_{k = 1}^K   {\bf{w}}_k^ns_k^n\left( t \right)\in{\mathbb C}^{L_{\rm T}\times1}\label{eqisacsignal},
\end{align}
where ${\bf w}_k^n\in{\mathbb C}^{L_{\rm T}\times1}$ is the downlink beamforming vector with power constraint $\sum\nolimits_{k = 1}^K {{{\left\| {{{\bf{w}}_k^n}} \right\|}^2}}  \le P$ and is operated on the information signal for the $k$-th user, i.e.,
\begin{align}
s_k^n\left( t \right) = \sum\limits_{m = 0}^{M_2^n - 1} {\sum\limits_{b = 0}^{B - 1} {\left[ {\tilde s_k^n\left[ {mb} \right]{e^{{\rm{j}}2\pi b{\Delta _f}\left( {t - {T_{cp}} - \left( {M_1^n + m} \right)T} \right)}}} \right.} }\nonumber \\
\left. { \times {\rm{rect}}\left( {t - \left( {M_1^n + m} \right)T} \right)} \right],M_1^nT \le t \le MT.
\end{align}
Here, $\tilde s_k^n\left[ {mb} \right]$ represents the complex modulated signal with power $\frac{1}{B}$ on the $b$-th subcarrier of the $m$-th OFDM symbol in Stage-II.
The parameters $B$, ${\Delta _f}=\frac{1}{T_o}$, ${T_{cp}}$, ${T_o}$, and $T=T_o+T_{cp}$ are defined as the number of subcarriers, subcarrier bandwidth, cyclic prefix duration, OFDM elementary symbol duration,  OFDM symbol duration including the cyclic prefix, respectively.
Besides, $M_1^n$ and  $M_2^n=M-M_1^n$ are the numbers of symbol slots allocated to Stage-I and Stage-II in the $n$-th frame, respectively.
Here, the rectangular pulse shape is applied for simplicity, i.e.,  ${\rm{rect}}(x)=1 $ if $0\le x\le T $, otherwise ${\rm{rect}}(x)=0 $ \cite{braun2014ofdm}.

In the following sections, we first derive the communication and radar performances when the CSI is updated through estimation or prediction, and then apply them to jointly optimize the CSI updating decisions and transmit beamforming metrics to reduce overall training costs and improve system utility.

\section{ Communication and Radar Performance Derivations}\label{sec3}
This section derives the communication and radar performances when the CSI updating decisions are given.
Firstly, we calculate the communication rate under predicted CSI or re-estimated CSI. Next, we present prediction and estimation methods for updating radar CSI and evaluate the radar performance using PCRB and SIC costs.

\subsection{ Communication Performance Derivation}
\subsubsection{Communication CSI Updating in Stage-I}
Let ${\bf{a}}_n^{\rm{c}} = \left[ {{{{a}}}_n^{\rm{c}}\left[ 1 \right], \cdots, {{{a}}}_n^{\rm{c}}\left[ K \right]} \right]^{ T} \in {{\mathbb C}^{K \times 1}}$ be the CSI updating decision vector for all communication users, where ${ {a}} _{n}^{\rm c}\left[ k \right] \in\left\{0,1\right\}$  denotes the binary CSI updating decision for user $k$ in the $n$-th frame, e.g., ${ {a}} _{n}^{\rm c}\left[ k \right]=0$ indicates that the CSI is predicted using temporal correlation in \eqref{eqtcorrC} with the last updated CSI, i.e., ${\bf{\hat g}}_k^{n-1}$, and ${ {a}} _{n}^{\rm c}\left[ k \right]=1$ indicates that the CSI is estimated by linear minimum mean squared error (MMSE) estimator \cite{papazafeiropoulos2016impact}. Then, the communication CSI of user $k$ in the $n$-th frame  is updated by
 \begin{align}\label{eqchannel}
 {\bf{\hat g}}_k^n = \left\{ {\begin{array}{*{20}{l}}
{{\rho _k}{\bf{\hat g}}_k^{n - 1},\;{\rm if}\;{ {a}} _{n}^{\rm c}\left[ k \right] = 0},\\
{ {\varkappa_k}\left( {\frac{1}{{DP_k^u}}{\bf{Y}}_k^n{{\bf{b}}_k}} \right),\;{\rm{if}}\;a_n^{\rm{c}}\left[ k \right] = 1}.
\end{array}} \right.\end{align}
Here, ${\bf{b}}_{k}\in{\mathbb C}^{ D\times 1}$  is the uplink training sequence transmitted by user $k$ with average power $P_k^{u}$
for each symbol,  and ${\bf{Y}}_{k}^n = {\bf{g}}_{k}^n{\bf{b}}_{k}^H + {\bf{U}}_{k}^{n}\in{\mathbb C}^{L_{\rm T}\times D}$ is the corresponding received sequences at the BS with Gaussian noise, i.e., ${\bf{U}}_k^n \sim {\cal CN}\left( {0,\delta {{\bf I}_{{L_{\rm T}}}}} \right)$. Also, we have ${\varkappa _k} = \frac{{D{\bar \beta _k}P_k^u}}{{D{\bar \beta _k}P_k^u + \delta }}$ from \cite{papazafeiropoulos2016impact}.

Then,   the estimation/prediction error of ${\bf g}_k^n$ is given by
  \begin{align}
 {\bf{e}}_k^n &= {\bf{g}}_k^n - {\bf{\hat g}}_k^n \nonumber\\
& = \left\{ {\begin{array}{*{20}{l}}
{{\rho _k}{\bf{e}}_k^{n - 1} + \sqrt {\left( {1 - \rho _k^2} \right)} {\bm{\varepsilon }}_k^n,\;{\rm if}\;{ {a}} _{n}^{\rm c}\left[ k \right]= 0},\\
\left( 1 - \varkappa _k\right)  {\bf{g}}_k^n- \frac{{{\varkappa _k}}}{{DP_k^u}}{\bf{U}}_k^n{{\bf{b}}_k},{\rm{if}}\;a_n^{\rm{c}}\left[ k \right]\; = 1.
\end{array}} \right.
 \end{align}
Next, from \cite{chen2023impact,papazafeiropoulos2016impact}, we know  ${\bf{e}}_k^n \sim {\cal CN}\left( {0,\varsigma _{{k}}^n{{\bf{I}}_{{L_{\rm T}}}}} \right)$, where
 \begin{align}
\varsigma _{{{k}}}^n = \left\{ {\begin{array}{*{20}{l}}
{\rho _k^2\varsigma _{{k }}^{n-1} + \left( {1 - \rho _k^2} \right){\bar\beta _k},\; {\rm if}\;{ {a}}_{n}^{\rm c}\left[ k \right]= 0,}\\
{{\bar\beta _k}\left( {1 - {\varkappa _k}} \right),\; {\rm if}\;{ {a}} _{n}^{\rm c}\left[ k \right] = 1}.
\end{array}} \right. \label{eqa8}
\end{align}

\subsubsection{Communication Rate in Stage-II}

With the assumption of frequency-flat channels, we only analyze the data rate on one subcarrier.
By transmitting the ISAC signals in \eqref{eqisacsignal} from the BS, the received signals on one subcarrier at user $k$ are
\begin{align}
y_{kn}^{\rm{c}}\left[ {mb} \right] = {\left( {{\bf{\hat g}}_k^n + {\bf{e}}_k^n} \right)^H} {\sum\nolimits_{k = 1}^K {{\bf{w}}_k^n} \tilde s_k^n\left[ {mb} \right]}  + u_{kn}^{\rm{c}}\left[ {mb} \right], \label{eq025}
\end{align}
where $ u_{kn}^{\rm{c}}\left[ {mb} \right]$ is the received complex Gaussian noise with power $\sigma_k$.
From \cite{omid2021low}, by considering the overhead of  pilot transmission in Stage-I, the overall effective transmission rate (nats/s/Hz) of user $k$ on each subcarrier in frame $n$ is
 \begin{align}
 C_k^n \!= \!\frac{{M - M_1^n}}{M}\log \left( {1 + \frac{{{{\left| {{{( {{\bf{\hat g}}_k^n} )}^H}{\bf{w}}_k^n} \right|}^2}}}{{\sum\limits_{i \ne k}^K {{{\left| {{{( {{\bf{\hat g}}_k^n} )}^H}{\bf{w}}_i^n} \right|}^2}}  + \varsigma _{k}^nP + {\sigma _k}}}} \right),\label{eq026}
\end{align}
where $M_1^n = D\sum\nolimits_{k = 1}^K {{\mathbb{I}}_{\left\{ {x\left| {x > 0} \right.} \right\}}\left( {{ {a}}_n^{\rm{c}}\left[ k \right]} \right)} $ is the time duration of Stage-I and also denotes the training overhead for CSI re-estimation of the selected communication users.

\subsection{ Radar Performance Derivation}

\subsubsection{Radar CSI Updating in Stage-II}
Let ${\bf{a}}_n^{\rm{r}} = \left[ {{{{a}}}_n^{\rm{r}}\left[ 1 \right], \cdots, {{{a}}}_n^{\rm{r}}\left[ Q \right]} \right]^T \in {{\mathbb C}^{Q \times 1}}$ be CSI updating decision vector for all radar targets, where ${ {a}} _{n}^{\rm r}\left[ q \right] \in\left\{0,1\right\}$  denotes  the binary CSI updating decision for  target  $q$ during Stage-II in the $n$-th frame, e.g., ${ a} _{n}^{\rm r}\left[ q \right]=0$ indicates that the CSI is predicted by leveraging temporal correlation in \eqref{eqtcorrR} with the last updated ${{\bf{\hat x}}_q^{n - 1}}$, i.e.,
\begin{align}
{\bf{\widetilde x}}_q^n \buildrel \Delta \over = \left[ {\widetilde \theta_q^n,\widetilde d_q^n,\widetilde v_q^n} \right]^{T}= \Gamma \left( {{\bf{\hat x}}_q^{n - 1}} \right)\in{\mathbb C}^{3\times1}.  \label{eqrp1}
\end{align}
Besides, ${ a} _{n}^{\rm r}\left[ q \right]=1$ indicates that the radar CSI of target $q$ is re-estimated from the echo signals using EKF algorithm \cite{ristic2003beyond}.

To perform radar CSI re-estimation,  the BS will receive the echoes from the target and the self-transmitted ISAC signals defined in \eqref{eqisacsignal}.
With the full-duplex operation, we follow a similar assumption in \cite{baghani2019dynamic,mei2022multi,chen2019exploiting} that  the self-interference is canceled by applying SIC, and the remaining echo signals at the BS for tracking are expressed as:
\begin{align}
{\bf{y}}_n^{\rm{r}}\left( t \right) = \sqrt {{L_{\rm{R}}}{L_{\rm{T}}}} \sum\nolimits_{q = 1}^Q {\left[ {\bar \alpha _q^n{e^{{\rm{j}}\phi _q^n}}{e^{{\rm{j}}2\pi \nu _q^nt}}{{\bf{v}}_{\rm{R}}}\left( {\theta _q^n} \right)} \right.} \nonumber\\
\left. {{\bf{v}}_{\rm{T}}^H\left( {\theta _q^n} \right){{\bf{s}}_n}\left( {t - \tau _q^n} \right)} \right] + {\bf{u}}_n^{\rm{r}}(t) \in{\mathbb C}^{L_{\rm R}\times1},\label{eq13}
\end{align}
where $ {\bf{u}}_n^{\rm{r}}(t)$ is received Gaussian noise with power density $\tilde \delta$. Here, $\bar\alpha _q^n = \sqrt {\frac{{c_0^2{\sigma _{{\rm{RCS}},q}}}}{{{{\left( {4\pi } \right)}^3}f_c^2{{\left( {d_q^n} \right)}^4}}}} $,
${\tau_q^n} = \frac{{2{d_q^n}}}{{{c_0}}}$, and
$\nu _q^n = \frac{{2{v_q^n}}\cos \left( {{\theta _q^n} - \bar \theta _q } \right)}{c_0}{f_c}$  are the magnitude attenuation, time delay, and Doppler phase shift of the $q$-th target in the $n$-th frame, respectively. Besides, $\phi _q^n$ is the corresponding phase noise and  ${\sigma _{{\rm{RCS,}}q}}$  is the complex radar cross-section (RCS) coefficient of target $q$. Then, $c_0$ and $f_c$ are the speed of light and subcarrier frequency, respectively.
Moreover, ${\bf{v}}_{\rm R}\left( {{\theta }} \right)$ and ${{\bf{v}}}_{\rm T}\left( {{\theta}} \right)$ are the receive and transmit steering
vectors with angle $\theta$, respectively. From \cite{chen2019channel}, assuming  half-wavelength antenna spacing,  we have $
{\bf{v}}_{\rm R}\left( {{\theta }} \right){\rm{ = }}\frac{1}{{\sqrt {{L_{\rm R}}} }}{\left[ {1,{e^{j\pi \sin \theta }},.,{e^{j\pi \left( {{L_{\rm R}} - 1} \right)\sin \theta }}} \right]^H} $ and $
{\bf{v}}_{\rm T}\left( {{\theta }} \right){\rm{ = }}\frac{1}{{\sqrt {{L_{\rm T}}} }}{\left[ {1,{e^{j\pi \sin \theta }},.,{e^{j\pi \left( {{L_{\rm T}} - 1} \right)\sin \theta }}} \right]^H} $, respectively.

Then, we have the following theorem to update the radar CSI and obtain the corresponding estimation performance.
\begin{theorem}\label{theorem1}
From \eqref{eqtcorrR} and  \eqref{eq13}, the predicted and posterior estimations of radar CSI ${\bf{\hat x}}_q^n$ using EKF method are given by
\begin{align}
\!\!\!{\bf{\hat x}}_q^n\! =\! \left\{\!\!\! \!{\begin{array}{*{20}{l}}
{\Gamma({\bf \hat x}_q^{n - 1}),\;{\rm if}\;{ a} _{n}^{\rm r}\left[ q \right] = 0},\\
\Gamma({\bf \hat x}_q^{n - 1}) + {\bf{K}}_q^n\left( {{\bf{\bar x}}_q^n - \Gamma({\bf \hat x}_q^{n - 1})} \right),\;{\rm if}\;{ a} _{n}^{\rm r}\left[ q \right] = 1,
\end{array}} \right.\label{eq19}
\end{align}
where ${\bf{\bar x}}_q^n $ is the direct estimation  of ${\bf x}_q^n$ defined in \eqref{eqstate}. Besides, we have $\gamma _{qn}^{\rm{r}} = {\sum\nolimits_{k = 1}^K {{{\left| {{\bf{v}}_{\rm{T}}^H\left( {\theta _q^n} \right){\bf{w}}_k^n} \right|}^2}} } $, ${\bm \Gamma} _q^{n - 1} = \frac{{\partial \Gamma \left( {\bf{x}} \right)}}{\partial{\bf{x}}}\left| {_{{\bf{x}} = {\bf{\hat x}}_q^{n - 1}}} \right.$,  and
\begin{align}
{\bf{K}}_q^n &= {\bf{\widetilde M}}_q^n( {\frac{1}{{{\gamma _{qn}^{\rm r}}}}{{\bm \Sigma} _{qn}^\delta}  + {\bf{\widetilde M}}_q^n} )^{ - 1}\in{\mathbb C}^{3\times3},\label{eqa16}\\
{\bf{\widetilde M}}_q^n &= {\bm{\Sigma }}_q^{\rm{\epsilon }} +
{{\bm{\Gamma}}_q^{n - 1}}{\bf{ M}}_q^{n - 1}({{\bm{\Gamma}}_q^{n - 1}})^H\in{\mathbb C}^{3\times3}.\label{eqb16}
\end{align}
Here, ${{\bm \Sigma} _{q}^\epsilon} $ and ${{\bm \Sigma} _{qn}^\delta} $  are defined in \eqref{eqtcorrR} and \eqref{eqstate}, respectively.

Besides, ${\bf{M}}_q^n\in{\mathbb C}^{3
\times3}$ is the corresponding PCRB matrix, i.e.,
\begin{align}&{\mathbb{E}}\left( {\left( {{\bf{\hat x}}_q^n - {\bf{x}}_q^n} \right){{\left( {{\bf{\hat x}}_q^n - {\bf{x}}_q^n} \right)}^H}} \right)\nonumber\\
\succeq & {\bf{M}}_q^n    \buildrel \Delta \over =  \left\{ {\begin{array}{*{20}{l}}
{{\bf{\widetilde M}}_q^n,\;{\rm if}\;{ a} _{n}^{\rm r}\left[ q \right]= 0,}\\
{{\bf{\bar M}}_q^n,\;{\rm if}\;{ a} _{n}^{\rm r}\left[ q \right]= 1,}
\end{array}} \right.\label{eq20}
\end{align}
where
${\bf{\bar M}}_q^n={\left( {{{( {{\bf{\widetilde M}}_q^n} )}^{ - 1}} + \gamma _{qn}^{\rm{r}}{{\left( {{\bf{\Sigma }}_{qn}^\delta } \right)}^{ - 1}}} \right)^{ - 1}} \in{\mathbb C}^{3
\times3}$.
\end{theorem}
\begin{IEEEproof}
Please refer to Appendix \ref{theorem1proof}
\end{IEEEproof}

\subsubsection{Weighted Radar Tracking Error and Cost in Stage-II}
Note that once the radar CSI ${\bf x}_q^n$  is determined to be re-estimated, the BS needs to perform SIC to cancel the self-interferences, and then we have \eqref{eq13} for radar CSI estimation. However, such SIC operation introduces extra complexity and energy costs, which are inversely proportional to the signal-to-interference-plus-noise ratio (SINR).
Therefore, this paper defines the radar performance as the weighted-sum of tracking errors (PCRBs ) and the energy costs of SIC.
Specifically,  the weighted PCRBs of $\theta _q^n$, $d _q^n$, and $v _q^n$ can be expressed as:
\begin{align}
{{\widetilde R}}_{qn}^{\rm Error} &= \sum\nolimits_{l = 1}^3 {{\omega _l}{\bf{M}}_q^n\left( {l,l} \right)},\label{eqtrackinerror1}
\end{align}
where ${\bf{M}}_q^n\left( {l,l} \right)$ is the $l$-th diagonal element of ${\bf M}_q^n$ in \eqref{eq20} and ${\omega _l}$ are non-negative constant weights to balance trade-offs among  $\theta _q^n$, $d _q^n$, and $v _q^n$.

With \eqref{eqa16}-\eqref{eq20}, we define ${{\bf{B}}_{qn}} = {\left( {{\bf{\Sigma }}_{qn}^\delta } \right)^{\frac{1}{2}}}{\left( {{\bf{\widetilde M}}_q^n} \right)^{ - 1}}{\left( {{\bf{\Sigma }}_{qn}^\delta } \right)^{\frac{1}{2}}}$ and denote its eigenvalue decomposition by ${{\bf{B}}_{qn}} = {{\bf{U}}_{qn}}{{\bf{\Lambda }}_{qn}}{\bf{U}}_{qn}^H$. Then, ${\bf{\bar M}}_q^n$ in \eqref{eq20} can be rewritten by
\begin{align}
{\bf{\bar M}}_q^n &= {\left( {{\bf{\Sigma }}_{qn}^\delta } \right)^{\frac{1}{2}}}{\left( { {{{\bf{B}}_{qn}} + \gamma _{qn}^{\rm{r}}} } \right)^{ - 1}}{\left( {{\bf{\Sigma }}_{qn}^\delta } \right)^{\frac{1}{2}}}\nonumber\\
&= {{\bf{C}}_{qn}}{\left( {{{\bf{\Lambda }}_{qn}} + \gamma _{qn}^{\rm{r}}} \right)^{ - 1}}{\bf{C}}_{qn}^H, \end{align}
where ${{\bf{C}}_{qn}} = {\left( {{\bf{\Sigma }}_{qn}^\delta } \right)^{\frac{1}{2}}}{{\bf{U}}_{qn}}$.
Next, ${{\widetilde R}}_{qn}^{\rm Error}$ in \eqref{eqtrackinerror1} is rewritten by
\begin{align}
{{\widetilde R}}_{qn}^{\rm Error} = { a} _{n}^{\rm r}\left[ q \right]\left( {\sum\limits_{l = 1}^3 {\frac{{\psi _{qn}^l}}{{\gamma _{qn}^{\rm{r}}+\lambda _{qn}^l }}} } \right) + \left( {1 - { a} _{n}^{\rm r}\left[ q \right]} \right)\tilde \psi _{qn},\label{eq22}
\end{align}
where ${\psi _{qn}^l = \sum\limits_{j = 1}^3 {\omega _j}{{{\left| {{\bf{C}}_{qn}^{jl}} \right|}^2}} }$ and $\tilde \psi _{qn} = \sum\limits_{j = 1}^3{\omega _j} {{\bf{\widetilde M}}_{qn}^{jj}}  $. Here, ${\bf{C}}_{qn}^{jl}$ and ${\bf{\widetilde M}}_{qn}^{jj}$ are the $(j,l)$-th and  $(j,j)$-th entries of ${\bf{C}}_{qn}$ and ${\bf{\widetilde M}}_{qn}$, respectively, and ${\lambda _{qn}^l}$ is the $l$-th diagonal  entry of ${{\bf{\Lambda }}_{qn}}$.

On the other hand, the costs of SIC can be mathematically expressed as  \cite{baghani2019dynamic,mei2022multi}
\begin{align}
{{\bar R}}_{qn}^{\rm Cost} = { a} _{n}^{\rm r}\left[ q \right]\left({{\xi _a} - {\xi _b}\lg  {\frac{{\gamma _{qn}^{\rm{r}}}}{{{\xi _c}P + {\sigma }}}} } \right),\label{eq23}
\end{align}
where ${\xi   _a} $ and ${\xi _b} $ are positive constant parameters denoting costs of SIC processing  \cite{mei2022multi}, while ${\xi _c} $ is the positive constant parameter related to the self-interference channel response. Note that there is no cost of SIC if the radar CSI is predicted.

Finally, using the derived tracking error (PCRB) in \eqref{eq22} and the costs of SIC in  \eqref{eq23}, the  weighted  radar tracking error and cost performance  can be expressed as
  \begin{align}
  {{R}}_q^n  = \bar \omega {{\widetilde R}}_{qn}^{\rm Error}  + \left( {1 - \bar \omega } \right){{\bar R}}_{qn}^{\rm Cost},\label{eq021a}
\end{align}
where $0\le{\bar \omega }\le1$ is the constant weight to balance tracking errors and costs.

\section{{Problem Formulation and Algorithm Development}}
This section formulates the problem of the joint intermittent CSI estimation and beamforming optimization and introduces a DROL framework to solve it sub-optimally.

\subsection{Problem Formulation}
This paper aims to jointly optimize the intermittent CSI updating decisions (i.e., ${\bf{a}}_n^{\rm{c}}$ and ${\bf{a}}_n^{\rm{r}}$) and downlink beamforming vectors (i.e., ${{\bf{W}}_n} = \left[ {{\bf{w}}_1^n, \cdots, {\bf{w}}_K^n} \right] \in {{\mathbb{C}}^{{L_{\rm T}} \times K}}$) to adapt to individual time-varying channel characteristics, thus maximizing the weighted communication achievable rates and minimizing the weighted radar tracking error and cost performances. Mathematically, we have the following system utility optimization problem in the $n$-th frame:
  \begin{align}
\mathop {{\rm{max}}}\limits_{{\bf{a}}_n^{\rm{c}},{\bf{a}}_n^{\rm{r}},{{\bf{W}}_n}} {\rm{ }}& {\cal U}\left( {{\bf{a}}_n^{\rm{c}},{\bf{a}}_n^{\rm{r}},{{\bf{W}}_n}} \right)  \buildrel \Delta \over =  \sum\limits_{k = 1}^K {\bar w_k^{\rm{c}}C_k^n}  - \sum\limits_{q = 1}^Q {\bar w_q^{\rm{r}}R_q^n} \tag{{\bf P1}} \label{aP1}\\
{\rm{s}}.{\rm{t}}.\;\;\;&   {\sum\nolimits_{k = 1}^K {\left\| {{\bf{w}}_k^n} \right\|}  \le {P}}  , \label{eqP1sj1} \\
 & a_n^{\rm{c}}\left[ k \right] \in \left\{ {0,1} \right\},1 \le k \le K, \label{eqP1sj3a}\\
 &a_n^{\rm{r}}\left[ q \right] \in \left\{ {0,1} \right\},1 \le q \le Q, \label{eqP1sj3}
   \end{align}
where $\bar w_k^{{\rm c}}$ and $\bar w_q^{{\rm r}}$ are non-negative constant weights to balance communication and tracking performances, respectively. Here, \eqref{eqP1sj1} is the maximum transmit power constraint at the BS, and
\eqref{eqP1sj3a} and \eqref{eqP1sj3} are binary CSI updating constraints to indicate whether the communication and radar CSI should be re-estimated or predicted.

\begin{figure*}[t]
\centering
{\center\includegraphics[width= 0.98\textwidth]{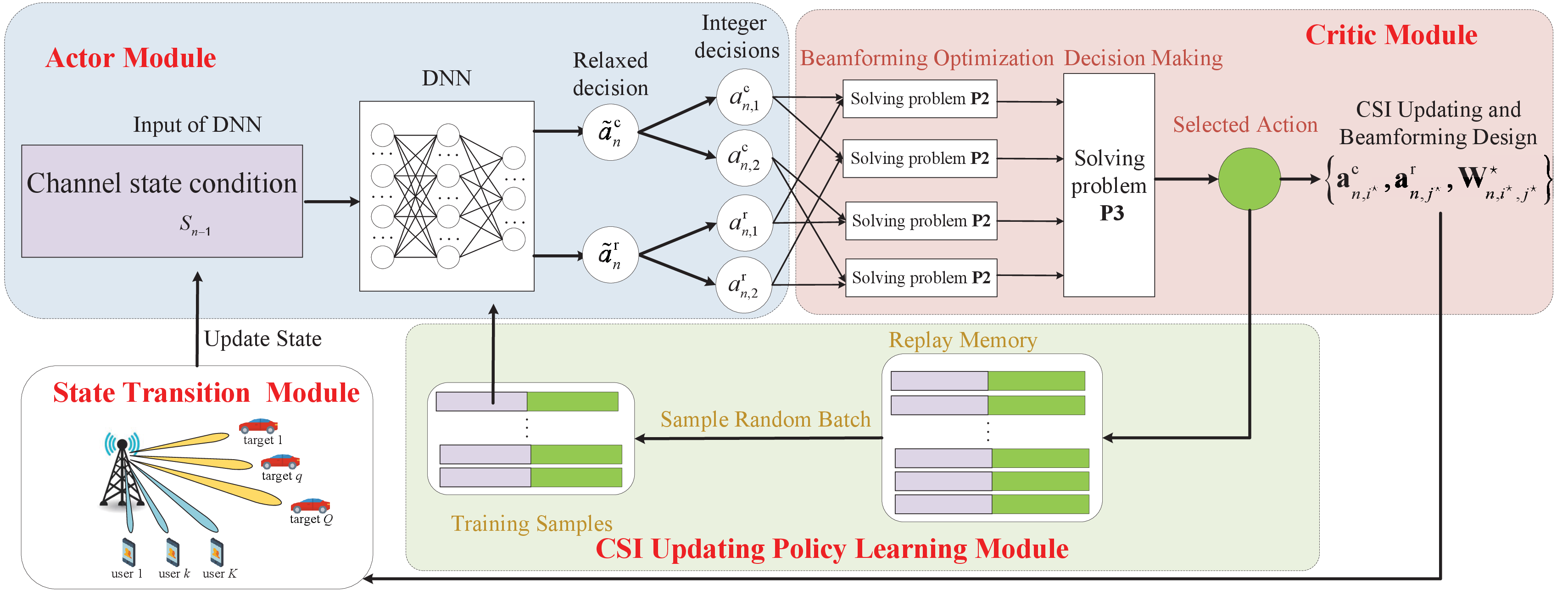} }
\caption{DROL-based dynamic intermittent CSI estimation framework together with transmit beamforming optimization. } \label{figDROL}
\end{figure*}
\subsection{Deep Reinforcement Online Learning (DROL) Framework}
Problem $\bf P1$ is unsolvable due to the causality issue.
Because optimizing CSI updating decisions requires both estimated and predicted CSI for performance comparisons. However, the CSI updating decisions are made in Stage-0, while the re-estimated CSI is obtained in Stage-I. As a result, the re-estimated CSI cannot be obtained during the optimization of the CSI updating decision in Stage-0.
Besides, $\bf P1$ is an MINLP problem, which is NP-hard in general and requires an exhaustive search to obtain the optimal solution \cite{chen2023WirelessService}. Hence, it can hardly be efficiently solved in real-time scenarios using conventional optimization algorithms.

Note that, if the binary CSI updating decisions ${\bf{a}}_n^{\rm{c}}$ and ${\bf{a}}_n^{\rm{r}}$ can be made before channel estimation, there is no causality issue and the remaining problem of $\bf P1$ no longer has integer constraints. Then, it can be efficiently solved by using convex optimization methods.
Therefore, we propose the following DROL framework, as shown in Fig. \ref{figDROL}, to solve problem $\bf P1$ in two main steps:
\begin{itemize}
\item Intermittent CSI estimation policy learning: We enable the DNN to perform the mapping from the channel conditions in the last frame to the CSI updating policy in the current frame by exploiting the experienced reinforced mapping. This step involves the actor, transition, learning, and critic modules in Fig. \ref{figDROL}, and the details will be shown in Section~V.
\item  Beamforming optimization: Based on the CSI updating decisions from DNN, we propose an efficient convex optimization algorithm to solve the remaining beamforming problem. This step is performed in the critic module in Fig. \ref{figDROL}, and the details will be shown in Section~VI.
\end{itemize}

The above DROL framework achieves CSI decision-making without requiring the current estimated CSI for performance comparison, thus it can avoid the causality issue.
Besides, the DROL framework makes decisions through the trained DNN, which only contributes a small portion of processing latency~\cite{huang2019deep}.
The beamforming matrices will be optimized by using the efficient low-complex algorithm proposed in  Section~\ref{secSolutino}.
Hence, the DROL method can solve the complexity issues without causing much latency.
Moreover, the DROL method learns the CSI updating policy in a dynamic wireless environment, which can adapt to the change of user/target channel distributions.
In the following two sections, we will show the detailed steps of  DROL in solving $\bf P1$.

\section{DROL-based Intermittent CSI Estimation}

This section develops the DROL framework to solve the joint intermittent CSI estimation and downlink beamforming optimization in problem $\bf P1$.
As shown in Fig. \ref{figDROL}, the DROL framework includes the following four modules:
\begin{itemize}
\item Actor module: it enables the DNN and the quantizer to generate multiple potential communication and radar CSI updating schemes, e.g., ${\bf{a}}_{n,i}^{\rm{c}}$ and ${\bf{a}}_{n,j}^{\rm{r}}$  from the last updated channel state condition.
 \item Critic module: it firstly solves the beamforming problem with each potential CSI updating scheme, i.e., $\left({\bf a}_{n,i}^{\rm c}, {\bf a}_{n,j}^{\rm r}\right)$ and selects the optimal decision with the optimized beamforming matrix, i.e., $\left( {{\bf{a}}_{n,{i^\star}}^{\rm{c}},{\bf{a}}_{n,{j^\star}}^{\rm{r}},{\bf{W}}_{n,{i^\star},{j^\star}}^ \star } \right)$. Then, it evaluates system performance and adaptively refines parameters to balance performance and complexity.
\item CSI updating policy learning module: it adds the optimized mapping ${{{\cal S}_{n - 1}} \to \left( {{\bf{a}}_{n,{i^ \star }}^{\rm{c}},{\bf{a}}_{n,{j^ \star }}^{\rm{r}}} \right)}$ into the memory and re-train the DNN for a better CSI updating policy
\item State transition module: it models the state transition from ${\cal S}_{n-1}$ to ${\cal S}_{n }$ based on the solution $\left( {{\bf{a}}_{n,{i^\star}}^{\rm{c}},{\bf{a}}_{n,{j^\star}}^{\rm{r}},{\bf{W}}_{n,{i^\star},{j^\star}}^ \star } \right)$.
\end{itemize}

\subsection{Actor Module}
The actor module consists of two parts, i.e., a DNN part to generate one relaxed solution and an integer action exploration part to quantize the relaxed solution into multiple potential feasible integer solutions for performance improvement.

Firstly, let the channel state condition in  frame $n$  be
  \begin{align}\label{eqchnnalcondition}
 {{\cal S}_n} = \left\{ {{\mathop{\rm Re}\nolimits} \left\{ {{\bf{\hat g}}_k^n} \right\},{\mathop{\rm Im}\nolimits} \left\{ {{\bf{\hat g}}_k^n} \right\},\varsigma _{{k}}^n,{\bf{\hat x}}_q^n,{\bf{M}}_q^n,\forall k,\forall q} \right\}. 
  \end{align}
Then, in the $n$-th frame, the DNN accepts the input of the last updated ${{\cal S}_{n-1}}$ to generate one relaxed continuous solution including $ {\bf \widetilde a}_n^{\rm c}\in {\mathbb C}^{K \times 1}$ and ${\bf \widetilde a}_n^{\rm r}\in {\mathbb C}^{Q \times 1}$, i.e.,
   \begin{align}\!\!\!
\pi_{\bm \xi} \left( {{{\cal S}_{n - 1}}} \right) =  \left( {{\bf{\widetilde a}}_n^{\rm{c}},{\bf{\widetilde a}}_n^{\rm{r}}} \right),
\end{align}
where  $\pi_{\bm \xi}(\cdot)$ and ${\bm \xi}$ denote the input-output mapping and the parameters of DNN, respectively.
Besides, the $k$-th and $q$-th components of ${\bf{\widetilde a}}_n^{\rm{c}}$ and ${\bf{\widetilde a}}_n^{\rm{r}}$ are continuous variables between 0 and 1, i.e.,  ${{\widetilde a}}_n^{\rm{c}}\left[k  \right] \in \left[ {0,1} \right]$ and ${{\widetilde a}}_n^{\rm{r}}\left[ q  \right] \in \left[ {0,1} \right]$.

Then, the integer action exploration part quantizes the relaxed decision ${\widetilde {\bf a}_n^{\rm{x}}}\in{\mathbb C}^{X\times 1}$  into multiple feasible integer decisions, where ${\rm x}\in\left\{{\rm c}, {\rm r}\right\}$ and $X=K$ if ${\rm x} ={\rm c}$, otherwise $X=Q$.
Here, the quantizer should be carefully designed for the exploration-exploitation trade-off, thus guaranteeing good convergence performance.
Heuristically, the integer solution is expected to be close to the continuous solution ${\widetilde {\bf a}_n^{\rm{x}}}$, which is obtained by using the trained DNN from prior mapping experience. Such exploitation of DNN can enhance convergence performance, but there is a risk of getting trapped in local optimal solutions.
Hence, we also need to explore individually separated integer solutions apart from ${\widetilde {\bf a}_n^{\rm{x}}}$. This exploration can potentially improve system utility performance ${\cal U}\left( {{\bf{a}}_n^{\rm{c}},{\bf{a}}_n^{\rm{r}},{{\bf{W}}_n}} \right)$, but it may result in a slower convergence due to involving new DNN mapping relationships.

Therefore, we propose the greedy noisy-order preserving quantization method to generate multiple potential feasible integer actions, which involves the following two steps:
\begin{itemize}
\item Firstly, we generate the first  $\widetilde {\cal K}_n^{\rm x}$ integer solutions $\left\{ {{\bf{a}}_{n,i}^{\rm{x}}\left| {1 \le i \le \widetilde{\cal{K}}_n^{\rm{x}}} \right.} \right\}$ by applying the order-preserving method  on ${\widetilde {\bf a}_n^{\rm{x}}}$. Please refer to Appendix \ref{orderproof} for details.
\item Then,  we generate additional $\widetilde {\cal K}_n^{\rm x}$ integer solutions with the probability ${\cal P}_n^{\rm x}$ by applying the same quantization method on the noisy ${\widetilde {\bf a}_n^{\rm{x}}}$. Specifically, we add a Gaussian noise $\widetilde {\bf u}_n\in{\cal CN}(0,{\bf I}_{X})$ on $\widetilde {\bf a}_n^{\rm x}$ and apply the Sigmoid function to keep the elements of noisy  $\widetilde {\bf a}_n^{\rm x}$ belonging to $\left[ {0,1} \right]$, i.e., ${\rm Sigmoid}(\widetilde {\bf a}_n^{\rm x}+{\bf u}_n)$. By further applying the above order preserving order algorithm in \eqref{eqfirst1} and \eqref{eqfirst2} on ${\rm Sigmoid}(\widetilde {\bf a}_n^{\rm x}+\widetilde{\bf u}_n)$, we can  obtain the additional $  \widetilde {\cal K}_n^{\rm x}$ integer solutions.
\end{itemize}

Therefore, given parameters $\widetilde {\cal K} _n^{\rm{x}}$ and ${\cal P}_n^{\rm x}$, the number of potential communication/radar CSI decisions explored is
\begin{align}\label{eqkn}
{\cal K}_n^{\rm x} = \left\{ {\begin{array}{*{20}{l}}
{\widetilde {\cal K} _n^{\rm{x}}, {\rm with}\; {\rm probability} \;1-{\cal P}_n^{\rm x}},\\
{2\widetilde {\cal K}_n^{\rm{x}}, {\rm with}\; {\rm probability}\; {\cal P}_n^{\rm x}}.
\end{array}} \right.
\end{align}
Here, ${\cal P}_n^{\rm x}$ can be regarded as the frame-varying probability of introducing more explorations, which can dynamically control the value of ${\cal K}_n^{\rm x}$ when $\widetilde {\cal K}_n^{\rm{x}}$ is given.  Since larger ${\cal P}_n^{\rm x}$ means more number of potential solutions explored, its value actually balances the system performance and complexity. The detail method to setup the values of $\widetilde {\cal K} _n^{\rm{x}}$ and ${\cal P}_n^{\rm x}$ will be discussed in Section \ref{secp}.

By applying the above greedy noisy-order preserving quantization method on  ${\widetilde {\bf a}_n^{\rm{x}}}$ for both  ${\rm x}={\rm c}$ and  ${\rm x}={\rm r}$, we can obtain ${\cal K}_n^{\rm c}$ and ${\cal K}_n^{\rm r}$ relaxed decisions of
$\widetilde {\bf a}_n^{\rm c}$ and $\widetilde {\bf a}_n^{\rm r}$, respectively, e.g., $\left\{ {{\bf{a}}_{n,i}^{\rm{c}}\left| {1 \le i \le {\cal K}_n^{\rm{c}}} \right.} \right\}$ and $\left\{ {{\bf{a}}_{n,j}^{\rm{r}}\left| {1 \le j \le {\cal K}_n^{\rm{r}}} \right.} \right\}$.
  Therefore, there are total ${\cal K}_n={\cal K}_n^{\rm c}\times {\cal K}_n^{\rm r} $ potential CSI updating schemes of $\left({\bf a}_{n,i}^{\rm c} ,{\bf a}_{n,j}^{\rm r}\right)$ that will be evaluated in the critic module.

\subsection{Critic Module}

The critic module selects the optimal solution and evaluates the system performance, and then adaptively refines algorithm parameters to balance algorithm performance and complexity.

\subsubsection{Optimal Solution Selection}
To begin with, the critic module solves the beamforming problem with each explored CSI updating decision $\left({\bf a}_{n, i}^{\rm c},{\bf a}_{n,j}^{\rm r}\right)$, i.e.,
\begin{align} {\bf{W}}_{n,i,j}^ \star  = \mathop {\arg \max }\limits_{\bf{W}} \;\;{\cal U}\left( {{\bf{a}}_{n,i}^{\rm{c}},{\bf{a}}_{n,j}^{\rm{r}},{\bf{W}}} \right) \;{\rm s.t.} \;\eqref{eqP1sj1}. \tag{\bf P2}
\end{align}
Then, it selects the optimal decision from ${\cal K}_n$ potential candidates, i.e.,
\begin{align} \left( {{\bf{a}}_{n,{i^\star}}^{\rm{c}},{\bf{a}}_{n,{j^\star}}^{\rm{r}},{\bf{W}}_{n,{i^\star},{j^\star}}^ \star } \right) \buildrel \Delta \over =\!\!\!\!\!\!\!\!\!\!\! \!\!\!\!\mathop {\arg \max }\limits_{\left\{ {{\bf{a}}_{n,i}^{\rm{c}},{\bf{a}}_{n,j}^{\rm{r}},{\bf{W}}_{n,i,j}^ \star ,\forall i,\forall j} \right\}}\!\!\!\! \!\!\!\!\!\!\!\!\!\!\!{\cal U}\left( {{\bf{a}}_{n,i}^{\rm{c}},{\bf{a}}_{n,j}^{\rm{r}},{\bf{W}}_{n,i,j}^ \star } \right)\tag{\bf P3} .
\end{align}
Here, the algorithm to solve $\bf P2$ is investigated in Section~\ref{secSolutino}, while $\bf P3$ is solved by comparing objective values directly.

\subsubsection{Performance Evaluation}
The critic module evaluates two categories of utility performances, namely practical and genie-aided system utilities, based on whether the training overhead of CSI exploration is involved.

Firstly,   there are $ {\cal }{\cal K}_n^{\rm c}$ potential communication CSI updating decisions, e.g., $\left\{ {{\bf{a}}_{n,i}^{\rm{c}}\left| {1 \le i \le {\cal K}_n^{\rm{c}}} \right.} \right\}$, in each of which arbitrary $a_{n,i}^{\rm{c}}\left[ k \right] = 1$ means that the CSI of user $k$ should be re-estimated in  Stage-I.
This also implies that the index set of the selected users for re-estimating CSI in practice should be the index set of non-zero elements in all decisions $\left\{ {{\bf{a}}_{n,i}^{\rm{c}}\left| {1 \le i \le {\cal K}_n^{\rm{c}}} \right.} \right\}$, which can be denoted as the practical communication CSI decision, i.e., ${\bf{a}}_n^{{\rm{c, practical}}}$, and its $k$-th entry is
$
{{a}}_n^{{\rm{c,practical}}}\left[ k \right] = {{\mathbb I}_{\left\{ {x\left| {x > 0} \right.} \right\}}}\left( {\sum\nolimits_{i = 1}^{{\cal K}_n^c} {a_{n,i}^{\rm{c}}\left[ k \right]} } \right)
$. Then, the corresponding training overhead is $M_1^{n,{\rm{practical}}} =D\sum\nolimits_{k = 1}^K {{{\mathbb I}_{\left\{ {x\left| {x > 0} \right.} \right\}}}\left( {a_n^{{\rm{c}},{\rm{practical}}}\left[ k \right]} \right)}  $. Given  $({\bf{a}}_n^{{\rm{c, practical}}},{\bf{a}}_{n,j^*}^{{\rm{r}}})$, we can evaluate the practical system utility by solving  $\bf  P2$, and the associated evaluated performance is called practical system utility.

However,  the final objective is to find the optimized CSI updating decision
$ ({\bf{a}}_{n,{i^ \star }}^{\rm{c}},{\bf{a}}_{n,{j^ \star }}^{\rm{r}})$ instead of  $({\bf{a}}_n^{{\rm{c, practical}}},{\bf{a}}_{n,j^*}^{{\rm{r}}})$.
To enable the DNN to learn the mapping from ${\cal S}_{n-1}$ to
$ ({\bf{a}}_{n,{i^ \star }}^{\rm{c}},{\bf{a}}_{n,{j^ \star }}^{\rm{r}})$, we need to do the following steps. Firstly, we define the genie-aided system utility by the optimal objective value of problem $\bf P3$, i.e.,  ${\cal U}\left( {{\bf{a}}_{n,{i^\star}}^{\rm{c}},{\bf{a}}_{n,{j^\star}}^{\rm{r}},{\bf{W}}_{n,{i^\star},{j^\star}}^ \star } \right)$.
Here, we only consider the training overhead for re-estimating the communication CSI of chosen users in ${\bf a}_{n, i}^{\rm c}$, i.e., $M_{1}^{n,\rm genie} = D\sum\nolimits_{k = 1}^K {{{\mathbb{I}}_{\left\{ {x\left| {x > 0} \right.} \right\}}}\left( {a_{n, i^\star}^{\rm{c}}\left[ k \right]} \right)} $, and ignore that in exploring the selected users in other communication CSI decisions $\left\{ {{\bf{a}}_{n,i}^{\rm{c}}\left| {1 \le i \le {\cal K}_n^{\rm{c}}},i \ne  i^\star \right.} \right\}$.
Then, we apply the genie-aided system utility instead of practical utility for DNN training, while we also evaluate the practical system utility that actually achieved in each frame, as shown in the simulations.

\subsubsection{Adaptive Parameter Refining} \label{secp}
The critic module needs to solve  $\bf P2$ in ${\cal K}_n$ times due to there are ${\cal K}_n={\cal K}_n^{\rm c}\times {\cal K}_n^{\rm r} $ potential CSI updating schemes. It is obvious that larger  ${\cal K}_n^{\rm r}$  brings better performance due to more potential  ${\bf a}_{n, i}^{\rm r}$ can be explored, although it also increases computational complexity.
Similarly,  a larger ${\cal K}_n^{\rm c}$ means more potential  ${\bf a}_{n, i}^{\rm c}$ will be explored and also means higher training overhead, thus it will not always increase the practical system utility.
To balance training overhead, complexity, and performance,  we apply the following intuitively adaptive procedure for setting  ${\cal K}_n^{\rm x}$, where ${\rm x}\in\left\{{\rm c}, {\rm r}\right\}$. The main idea is that we aim at small $\widetilde {\cal K}_n^{\rm x}$ and ${\cal P}_n^x$ in \eqref{eqkn} when the DNN converges, and it is sufficient to find the optimal decision with a small distance to ${\widetilde {\bf a}_n^{\rm{x}}}$. Heuristically, we propose the following refining method, i.e.,
\begin{align}
&\widetilde{\cal K} _n^{\rm{x}} = \left\{ {\begin{array}{*{20}{l}}
{\left\lfloor \frac{\sum\limits_{l=n - \widetilde \Delta^{\rm x}+1 }^n  {{\rm{mod}}\left( {I_l^x,X} \right)}}{{\widetilde \Delta^{\rm x}}}   \right\rfloor  + 1,{\rm{if}}\;{\rm{mod}}\left( {n,\widetilde \Delta^{\rm x} } \right) = 0},\\
{\widetilde {\cal K}_{n - 1}^{\rm{x}},{\rm otherwise}},
\end{array}} \right. \nonumber \\
&
{\cal P}_n^{\rm x} = 1-\frac{{\cal A }^{\rm{x}}}{{\widetilde {\cal K}_n^{\rm{x}}}},\label{equpdatinngpa1}
\end{align}
where ${\cal A }^{\rm{x}}$ is a constant parameter subject to ${\cal A} ^{\rm{x}} \le \mathop {\min }\limits_n \widetilde {\cal K}_n^{\rm{x}}$,  ${I_l^x}$ denotes the index of the  best solution ${\bf a}_{l,i^\star}^{\rm x}$ of solving  $\bf P3$ in the $l$-th frame, and $\widetilde \Delta^{\rm{x}}$ is the updating interval of ${\cal K}_{n - 1}^{\rm{x}}$. Besides, ${\rm mod}\left(x,y\right)$ is the modulo operator, which calculates the remainder when integer $x$ is divided by integer $y$.

\begin{algorithm}[t]
\caption{DROL based intermittent CSI updating policy learning and beamforming design }\label{algorithmP0}
{{
\begin{algorithmic}[1]

\FOR {$n=1:1:N$  }
 \STATE {\!\!\!\!\!\!\bf Actor Module:}{\\
Generate a relaxed decision of $\left( {\bf a}_n^{\rm c},{\bf a}_n^{\rm r}\right)$ by input  ${{\cal S}_{n-1}}$ into DNN, i.e., $\pi_{\bm \xi} \left( {{{\cal S}_{n - 1}}} \right) = \left( {{\bf{\widetilde a}}_n^{\rm{c}},{\bf{\widetilde a}}_n^{\rm{r}}} \right)$}\\
For ${\rm x}\in\left\{{\rm c}, {\rm r}\right\}$:  generate first $\widetilde {\cal K}_n^{\rm x} $ integer decisions ${\bf a}_{n,i}^{\rm x} $ using order-preserving quantization on ${\bf \widetilde a}_n^{\rm x}$, i.e., \eqref{eqfirst1} and \eqref{eqfirst2}, and set $ {\cal K}_n^{\rm x} =\widetilde {\cal K}_n^{\rm x} $\\
For ${\rm x}\in\left\{{\rm c}, {\rm r}\right\}$: { if} { ${\rm rand}()<{\cal P}_n^{\rm x}$}, then generate remaining $\widetilde {\cal K}_n^{\rm x} $ integer decisions ${\bf a}_{n,i}^{\rm x} $ using order-preserving quantization method on ${\rm Sigmoid}(\widetilde {\bf a}_n^{\rm x}+{\bf u}_n)$, and set $ {\cal K}_n^{\rm x} =2\widetilde {\cal K}_n^{\rm x} $ \\
 \STATE {\!\!\!\!\!\!\bf Critic Module:}{\\
 For $\forall i, j$, solve problem $\bf P2$ with constant $\left({\bf a}_{n,i}^{\rm c} ,{\bf a}_{n,j}^{\rm r}\right)$  using Algorithm \ref{algorithmP10} in Section VI and obtain ${\bf{W}}_{n,i,j}^ \star$\\
Update ${\cal P}_n^{\rm c}$, ${\cal P}_n^{\rm r}$, ${\widetilde{ \cal K}}_n^{\rm c}$, and ${\widetilde{ \cal K}}_n^{\rm r}$ using \eqref{equpdatinngpa1}\\
Solve $\bf P3$ to find the optimal $\left( {{\bf{a}}_{n,{i^\star}}^{\rm{c}},{\bf{a}}_{n,{j^\star}}^{\rm{r}},{\bf{W}}_{n,{i^\star},{j^\star}}^ \star } \right)$ from ${\cal K}_n$ potential integer solutions}
\STATE {\!\!\!\!\!\!\bf CSI Updating Policy Learning Module:}{\\
Apply ADAM algorithm to update DNN parameters ${\bm \xi}$ by reducing ${\cal L}\left( \bm \xi  \right)$ in \eqref{eqcrossentropy}\\ }
\STATE {\!\!\!\!\!\!\bf  State Transition Module:}{\\
Update   ${\bf{\hat g}}_k^n$ and $\varsigma _{{k}}^n$ by substituting ${\bf{a}}_{n,{i^\star}}^{\rm{c}}$ into  \eqref{eqchannel} and \eqref{eqa8}\\
Update
$\sigma _{\theta _q^n}^2$,$\sigma _{d _q^n}^2$ , and $\sigma _{v_q^n}^2$ in \eqref{escrlb} using ${\bf{W}}_{n,{i^\star},{j^\star}}^ \star   $\\
Update ${\bf{\hat x}}_q^n$  and  ${\bf{M}}_q^n$ by substituting ${\bf{a}}_{n,{j^\star}}^{\rm{r}}$ into  \eqref{eq19} and \eqref{eq20}\\
Update channel condition ${\cal S}_n$  in \eqref{eqchnnalcondition}
}
\ENDFOR
\end{algorithmic}}}
\end{algorithm}

\subsection{CSI Updating Policy Learning Module} 
This module trains DNN parameters ${\bm \xi}$  to achieve a better CSI updating policy by using the new optimized mapping sample, i.e.,  ${{{\cal S}_{n - 1}} \to \left( {{\bf{a}}_{n,{i^ \star }}^{\rm{c}},{\bf{a}}_{n,{j^ \star }}^{\rm{r}}} \right)}$  obtained from critic module in frame $n$. To do so, we first establish the replay memory to collect the most recent samples. Then, we train the DNN every frame and randomly
select a batch of samples $\left\{ {{{\cal S}_{l - 1}} \to \left( {{\bf{a}}_{l,{i^ \star }}^{\rm{c}},{\bf{a}}_{l,{j^ \star }}^{\rm{r}}} \right)\left| {l \in {{\mathbb L}_n}} \right.} \right\}$ from the replay memory, where ${{{\mathbb L}_n}} $ denotes the set of frame indices of training samples.

Then, parameter $\bm \xi $ is updated under the principle of reducing averaged cross-entropy loss using the adaptive moment estimation (ADAM) algorithm \cite{huang2019deep,yu2024phase}, i.e.,
\begin{align}
{\cal L}\left( \bm \xi  \right) =  - \frac{1}{{\left| {{{\mathbb L}_n}} \right|}}\sum\nolimits_{l \in {{\mathbb L}_n}} \left[ {{\left( {{\bf{a}}^\star_{{l}}} \right)}^T}\log {\pi _{\bm{\xi }}}\left( {{{\cal S}_{l - 1}}} \right) \right.\nonumber\\
\left.+ {{\left( {1 -  {{\bf{a}}^\star_{{l }}}} \right)}^T}\left( {1 - \log {\pi _{\bm{\xi }}}\left( {{{\cal S}_{l - 1}}} \right)} \right) \right]\label{eqcrossentropy},
\end{align}
where ${{\bf{a}}^\star_{{l}}}=\left[({{\bf{a}}_{l,{i^ \star }}^{\rm{c}})^T,({\bf{a}}_{l,{j^ \star }}^{\rm{r}}})^T\right]^T\in{ \mathbb C}^{(K+Q)\times1}$ and $\left| {{{\mathbb L}_n}} \right|$ denotes the number of samples involved for training.

\subsection{State Transition Module}

This module implements the state transition from ${\cal S}_{n-1}$ to ${\cal S}_{n}$ based on $ \left( {{\bf{a}}_{{{n,i}^ \star }}^{\rm{c}},{\bf{a}}_{{{n,j}^ \star }}^{\rm{r}},{{\bf{W}}^ \star }} \right)$.
For communication CSI updating,   ${\bf{\hat g}}_k^n$ and $\varsigma _{{k}}^n$ are updated by substituting ${\bf{a}}_{{{n,j}^ \star }}^{\rm{c}}$ into  \eqref{eqchannel} and \eqref{eqa8}, respectively. As for radar CSI updating, we first calculate the CRBs of $\theta _q^n$, $d _q^n$, and $v _q^n$
in \eqref{escrlb}, and then update ${\bf{\hat x}}_q^n$  and  ${\bf{M}}_q^n$ by substituting ${\bf{a}}_{{{n,j}^ \star }}^{\rm{r}}$ into  \eqref{eq19} and \eqref{eq20} and with the results in \eqref{escrlb}.
Finally, the  DROL for intermittent CSI estimation and beamforming design is summarized in Algorithm \ref{algorithmP0}

\section{ Solution to Beamforming Optimization}\label{secSolutino}
This section proposes an efficient algorithm to solve the beamforming optimization problem $\bf P2$ sub-optimally.

\subsection{Problem Transformation}
We transform the non-convex problem  ${\bf P2}$ into a form that can be solved through efficient optimization algorithms.

For the sake of simplicity, we will omit index $n$ where it does not lead to ambiguity. Then, given $\left({\bf a}_{n,i}^{\rm c} ,{\bf a}_{n,j}^{\rm r}\right)$  in  problem ${\bf P2}$, we can substitute $ {\bf a}_{n,i}^{\rm c}$  into \eqref{eqchannel} and \eqref{eqa8} to obtain $\varsigma _{{k}}^n $ and $ {\bf{\hat g}}_k^n $, as defined in \eqref{eq026}, respectively.
Then, upon  denoting  $
{A_k}\left( {\bf{W}} \right) = ({\bf{\hat g}}_{k }^n)^H{\bf{w}}_k^n $ and $
{B_k}\left( {\bf{W}} \right) = \sum\nolimits_{i = 1}^K {{{\left| {({\bf{\hat g}}_{k }^n)^H{\bf{w}}_{in}^{\rm{c}}} \right|}^2}}  + \sigma _{kn}^{\rm{c}}$ with $ \sigma _{kn}^{\rm{c}}= \varsigma _{k}^nP + {\sigma _k}$, and  $\bar w_{kn}^c= \bar w_{k}^cM_2^n/M$,  the weighted communication performance  is represented by
 \begin{align}
\sum\nolimits_{k = 1}^K {\bar w_k^{\rm{c}}C_k^n} & \!=\! \sum\nolimits_{k = 1}^K {\bar w_{kn}^{\rm c}\log \left( {1 \!+ \!\frac{{{{\left| {{A_k}\left( {\bf{W}} \right)} \right|}^2}}}{{B_k \left( {\bf{W}} \right) - {\left| {{A_k}\left( {\bf{W}} \right)} \right|^2}}}} \right)}\nonumber\\
& \buildrel \Delta \over = {{\cal F}_C}\left( {\bf{W}} \right).\label{eqa31}
\end{align}

Then, let ${\mathbb Q} = \left\{ {q\left| {a_{n,j}^{\rm{r}}\left[ q \right] = 1},1\le q \le Q \right.} \right\}$ be the index set of the selected radar targets whose CSI need to be re-estimated. Also, we substitute ${\bf a}_{n,j}^{\rm r}$ into \eqref{eq22} and \eqref{eq23} to calculate $
{{\widetilde R}}_{qn}^{\rm Error} $ and ${{\bar R}}_{qn}^{\rm Cost} $, respectively. From \eqref{eq021a},  the weighted radar tracking error and cost performance is represented by
 \begin{align}
 &\sum\nolimits_{q = 1}^Q {\bar w_q^{\rm{r}}R_q^n}  = \bar \Psi  \nonumber  \\
  &+ \underbrace {\sum\limits_{q \in {\mathbb Q}} {\bar w_q^{\rm{r}}\left( {\sum\limits_{l = 1}^3 {\frac{{\bar \omega \psi _{qn}^l}}{{{\mu _{q}} + \lambda _{qn}^l}} - \left( {1 - \bar \omega } \right){\zeta _b}\lg \left( {{\mu _{q}}} \right)} } \right)} }_{{{\cal F}_R}\left( {\bm{\mu }} \right)},\label{eqa32}
\end{align}
if ${\mu _{q}} = \gamma _{qn}^r$, where  $\gamma _{qn}^r \approx  {\sum\nolimits_{k = 1}^K {{{\left| {{\bf{v}}_{\rm{T}}^H\left( {{\widetilde \theta} _q^n} \right){\bf{w}}_k^n} \right|}^2}} }$ due to \eqref{eq19}, and ${\widetilde\theta _q^n}$ is defined in \eqref{eqrp1}. Besides, ${\bm \mu} \in  {\mathbb C}^{\left|{\mathbb Q}\right|\times1} $ is the vector consisting of all  auxiliary variables $\mu _{q}$ for $q\in\mathbb Q$, and the constant term $\bar \Psi$ is from the calculation of $\sum\nolimits_{q = 1}^{ Q } {\bar w_q^{\rm{r}}R_q^n}$ that is not related to $\bf W$.

Then, with the introduced auxiliary variables ${\bm \mu}$ and substituting \eqref{eqa31} and \eqref{eqa32} into  ${\bf P2}$, we have its equivalent problem:
 \begin{align}\label{eqproblem4}
 \mathop {{\rm{max}}}\limits_{{\bf{W}},{\bm{\mu }}}& \;\;{{\cal F}_C}\left( {\bf{W}} \right) - {{\cal F}_R}\left( {\bm{\mu }} \right)\tag{\bf P4}  \\
{\rm s.t.}\; &{\mu_q} \le \sum\nolimits_{k = 1}^K
{\left| {C_{kq}^n\left( {\bf{W}} \right)} \right|^2}, q\in{\mathbb Q}, \label{eq39} \\
&\eqref{eqP1sj1},\nonumber
\end{align}
where $C_{kq}^n\left( \bf W \right) = {\bf{v}}_{\rm{T}}^H\left( {\widetilde\theta _q^n} \right){\bf{w}}_k^n
$.

However, problem ${\bf P4}$ is challenging to solve because of the non-convexities present  in ${{\cal F}_C}\left( {\bf{W}} \right)$ and \eqref{eq39}.
Hence, we employ the FP and SCA schemes to handle the non-convexities in ${{\cal F}_C}\left( {\bf{W}} \right)$ and \eqref{eq39}, respectively.
Specifically, since ${{\cal F}_C}\left( {\bf{W}} \right)$ follows the form of the sum-of-logarithms-of-ratio problem, it can be transformed into the tractable form by applying the FP approach \cite{guo2020weighted}, i.e.,
\begin{align}
  {\cal F}_C \left( {\bf{W}} \right)\nonumber
=& \mathop {\max }\limits_{ {\bm{\alpha }}} \sum\nolimits_{k = 1}^K {\bar w_{kn}^{\rm c}\left( {\Omega \left( {{\alpha _k}} \right) + \left( {1 + {\alpha _k}} \right)\frac{{{{\left| {{A_k}\left( {\bf{W}} \right)} \right|}^2}}}{{{B_k}\left( {\bf{W}} \right)}}} \right)}\nonumber \\
=&\mathop {\max }\limits_{{\bm\alpha}, {\bm \beta} } \sum\nolimits_{k = 1}^K {\left( {\bar w_{kn}^{\rm{c}}\Omega \left( {{\alpha _k}} \right) - {{\left| {{\beta _k}} \right|}^2}{B_k}\left( {\bf{W}} \right)} \right.} \nonumber\\
&\qquad\qquad\left. { + 2\sqrt {\bar w_{kn}^{\rm{c}}\left( {1 + {\alpha _k}} \right)} {\rm{Re}}\left\{ {\beta _k^*{A_k}\left( {\bf{W}} \right)} \right\}} \right)
 \nonumber\\
=& \mathop {\max }\limits_{ {\bm{\alpha }},{\bm{\beta }}}  \widetilde{\cal  F}_C\left({{\bf{W}},{\bm{\alpha }},{\bm{\beta }}}\right),
\end{align}
where ${\bm \alpha}  = \left[ {{\alpha _1},\cdots,{\alpha _K}} \right]\in  {\mathbb C}^{K\times1}$ and  ${\bm \beta}  = \left[ {{\beta _1},\cdots,{\beta _K}} \right]\in  {\mathbb C}^{K\times1}$ are the corresponding auxiliary variables, and $\Omega \left( {{\alpha _k}} \right){\rm{ = }}\log \left( {1 + {\alpha _k}} \right) - {\alpha _k}$.
Here,  $\widetilde{\cal  F}_C\left({{\bf{W}},{\bm{\alpha }},{\bm{\beta }}}\right)$  is concave with respect to each of its variables ${\bf W}$, ${\bm\alpha}$, and ${\bm\beta}$  when other variables are fixed, e.g., if  ${\bm\alpha}$ and ${\bm\beta}$ are given,  $ \widetilde{\cal  F}_C\left({{\bf{W}},{\bm{\alpha }},{\bm{\beta }}}\right)$ is concave with respect to ${\bf W}$; if  ${\bf W}$ and ${\bm\beta}$ are given,   it is concave with respect to ${\bm\alpha}$; and if ${\bf W}$ and ${\bm\alpha}$ are given,   it is concave with respect to ${\bm\beta}$.

To address the non-convex constraint in \eqref{eq39}, we use the SCA method to iteratively approximate the convex function $\sum\nolimits_{k = 1}^K{\left| {C_{kq}^n\left( {\bf{W}} \right)} \right|^2}$  using its lower bound in an affine form, which is obtained by applying linear interpolation between ${\bf W}$ and ${{{\bf{W}}^{{\rm{old}}}}}$ ( the updated solution in the last iteration), i.e.,
 \begin{align}
{\left| {C_{kq}^n\left( {\bf{W}} \right)} \right|^2} & \ge 2{\mathop{\rm Re}\nolimits} \left( {C_{kq}^n{{\left( {{{\bf{W}}^{{\rm{old}}}}} \right)}^*}C_{kq}^n\left( {\bf{W}} \right)} \right) - {\left| {C_{kq}^n\left( {{{\bf{W}}^{{\rm{old}}}}} \right)} \right|^2}\nonumber\\
 &= C_{kq}^{{\rm{Lb}}}\left( {\bf{W}} \right).\label{eqsca1a}
\end{align}
Then, problem  ${\bf P4}$ can be transformed into
\begin{align}\label{eqproblem5}
 \mathop {{\rm{max}}}\limits_{{\bf{W}},{\bm{\mu}}, {\bm{\alpha }},{\bm{\beta }}}& { \widetilde{\cal  F}_C}\left( {{\bf{W}},{\bm{\alpha }},{\bm{\beta }}} \right)   - {{\cal F}_R}\left( {\bm{\mu }} \right)\tag{\bf P5} \\
{\rm{s}}.{\rm{t}}.\; & {\mu_q}\le \sum\nolimits_{k = 1}^K  C_{kq}^{{\rm{Lb}}}\left( {\bf{W}} \right), q\in{\mathbb Q},\label{eqsca1}\\
&\eqref{eqP1sj1}.\nonumber
\end{align}
Finally, this problem can be solved by alternating optimization, because it belongs to convex problems when optimizing each of ${\bf W}$, ${\bm\alpha}$,  ${\bm\beta}$, and $\bm \mu$ given other variables.

\subsection{Solution to Problem {\bf P5}}
This part presents alternating optimization to solve  ${\bf P5}$ with closed-form solutions. To begin with, the Lagrangian dual problem is given by
   \begin{align}
 \mathop {{\rm{min}}}\limits_{{\bm \eta}\ge0} \mathop {{\rm{max}}}\limits_{{\bf{W}},{\bm{\mu }},{\bm{\alpha }},{\bm{\beta }}} \;\; {\cal F}\left( {{\bf{W}},{\bm{\mu }},{\bm{\alpha }},{\bm{\beta }},{\bm \eta}} \right)  \;\;
{\rm{s}}.{\rm{t}}. \;\;\eqref{eqP1sj1},\tag{\bf P6}
\end{align}
where
\begin{align}  {\cal F}\left( {{\bf{W}},{\bm{\mu }},{\bm{\alpha }},{\bm{\beta }},{\bm \eta}} \right)  \buildrel \Delta \over =&  {{\widetilde { \cal F}}_C}\left( {{\bf{W}},{\bm{\alpha }},{\bm{\beta }}} \right)\! -\! {{\cal F}_R}\left( {\bm{\mu }} \right) \nonumber\\
  &+\sum\nolimits_{q \in {\mathbb Q}}  {{\eta _q}} \left( {\sum\nolimits_{k = 1}^K {C_{kq}^{{\rm{Lb}}}( {\bf{W}} )}  \!- \!{\mu_q}} \right).\nonumber
\end{align}
Here,
${\bm \eta} \in  {\mathbb C}^{\left|{\mathbb Q}\right|\times1} $ is the vector consisting of   $\eta _q$ for $q\in\mathbb Q$, and ${\eta _q}$ is the non-negative Lagrangian dual variable due to constraint \eqref{eqsca1}, which is updated by sub-gradient method \cite{yu2006dual}.

Let ${\bf{W}}^{\rm old}$, ${\bm{\mu }}^{\rm old}$, ${\bm{\alpha }}^{\rm old}$, ${\bm{\beta }}^{\rm old}$, and ${\bm \eta }^{\rm old}$ be the updated solutions in the last iteration. Then, due to the concavity of ${\cal F}\left( {\bm{\alpha }},{\bm{\beta }},{\bm{\mu }},{\bf{W}},{\bm \eta} \right)$ with respect to ${\bm{\alpha }}$, ${\bm{\beta }}$, ${\bm{\mu }}$, ${\bf{W}}$, respectively, by  equating the first derivative of  ${\cal F}\left(  {\bm{\alpha }},{\bm{\beta }},{\bm{\mu }},{\bf{W}},{\bm \eta} \right)$ in terms of ${\bm{\alpha }}$, ${\bm{\beta }}$, and ${\bm{\mu }}$ to zero successively,  we have
\begin{align}
\beta _k^{\rm new} &= \frac{{\sqrt {\bar w_{kn}^{\rm c}\left( {1 + \alpha _k^{\rm old}} \right)} {A_k}\left( {{{\bf{W}}^{\rm old}}} \right)}}{{{B_k}\left( {{{\bf{W}}^{\rm old}}} \right)}},\label{eqwoptimize1}\\
\alpha _k^{\rm new}&= \frac{{\mathchar'26\mkern-10mu\lambda _k^2 + {\mathchar'26\mkern-10mu\lambda _k}\sqrt {\mathchar'26\mkern-10mu\lambda _k^2 + 4} }}{2},\label{eqwoptimize0}\\
\mu _q^{\rm new}&= \mu _q^{ \dag }, q \in {\mathbb Q},\label{eqwoptimize2}
\end{align}
where ${\mathchar'26\mkern-10mu\lambda _k}=\frac{1}{{\sqrt {\bar w_{kn}^{\rm c}} }}{\mathop{\rm Re}\nolimits} \left\{ {{{\left( {\beta _k^{\rm new}} \right)}^*}({\bf{\hat g}}_{k }^n)^H{{{\bf{w}}}_k^{\rm old}} } \right\}$ and $\mu _q^{ \dag }$ is the solution to the following equation,
\begin{align}
\sum\limits_{l = 1}^3 \frac{{\bar\omega }\psi _{qn}^l}{{{( {\mu _q^{{ \dag }} + \lambda _{qn}^l} )}^2}}  + \frac{{\left( {1 - \bar\omega} \right){\zeta _b}}}{{\mu _q^{{ \dag }}\ln 10}} - \eta _q^{{\rm{old}}} = 0.
\end{align}
By applying Lagrangian method to optimize $\bf W$, we have
\begin{align}
{\bf{w}}_k^{{\rm{new}}} &= {\left( {\tilde \lambda {\bf{I}} + \sum\nolimits_{j = 1}^K {{{\left| {\beta _j^{{\rm{new}}}} \right|}^2}{{{\bf{\hat g}}}_{j}^n}({\bf{\hat g}}_{j}^n)^H} } \right)^{ - 1}}{{\bf{h}}_{k}^n},\label{eqwoptimize3}
\end{align}
where
\begin{align}{{\bf{h}}_{k}^n} =& \sqrt {\bar w_{kn}^{\rm{c}}\left( {1 + \alpha _k^{{\rm{new}}}} \right)} \beta _k^{{\rm{new}}}{{{\bf{\hat g}}}_{k}^n}\nonumber\\
& + \sum\nolimits_{q \in {\mathbb Q}} {\eta _q^{{\rm{old}}}{C_{kq}^{\rm Lb}}\left( {{{\bf{W}}^{{\rm{old}}}}} \right){\bf{v}}\left( {{\widetilde \theta _q^n}} \right)}.
\end{align}
Note that  $\tilde \lambda $ in \eqref{eqwoptimize3} is the optimal dual variable due to transmit power constraint from \eqref{eqP1sj1}, which can be obtained by using a linear search to make the equality hold in \eqref{eqP1sj1}. However,
calculating  \eqref{eqwoptimize3} requires matrix inverse operation, which may cause high complexity.
Therefore,  we can apply the prox-linear method to update $\bf W$ for further complexity reduction.  Please refer to Appendix \ref{appendixW} for details.

\begin{table*}[t]
 \centering
 \begin{tabular}{|c|c||c|c|}
\hline
{Parameters}      &{Value}                                  &{Parameters}            &{Value}\\ \hline
{Speed of light}     &{$c_0$ = $3\times10^8$ m/s }                      &{Number of subcarriers} &{$B$ = 64}\\ \hline
{Total signal bandwidth} &{$B\Delta_f$ = 10 MHz }&{Subcarrier bandwidth}&{$\Delta_f$ = 156.25 KHz} \\ \hline
 {Elementary OFDM symbol duration} &{$T_o$ = $1/\Delta_f$ = 6.4 us}  &{Cyclic prefix duration}      &{$T_{cp}$ = $\frac{1}{4}T$ = 1.6 us} \\ \hline
  Transmit OFDM symbol duration&{$T$ =  8 us}     &{Number of symbols }          &$M$ = 800\\ \hline
 Communication performance weight & $\bar w_k^{{\rm c}}=0.3$ & Radar performance weight & $\bar w_q^{{\rm r}}=20\times(1-0.3)$  \\ \hline
 Radar tracking error and cost weight &${\bar \omega }=0.3$ & SIC cost parameters & $\xi_a=0.5$, $\xi_b=0.24$, $\xi_c=1$  \\ \hline
 \end{tabular}
\caption{Simulation parameter setup}\label{table1}
\end{table*}
Next, we apply the sub-gradient method to update $\bm \eta$, i.e.,
  \begin{align}
\eta _q^{\rm new}\!=\! {\rm max} \left\{ {\eta _q^{\rm old} \!-\! {\Delta _s^{\bar t}}\left( {\sum\limits_{k = 1}^K {C_{kq}^{{\rm{Lb}}}\left( {\bf{W}}^{\rm new} \right)} \! -\! {\mu_q^{\rm new}}  } \right)},0 \right\},\label{eqwoptimize4}
\end{align}
where ${\Delta _s^{\bar t}}=\frac{\Delta _s}{\sqrt{\bar t}}$ represents the updated step size, where $\Delta _s$ is the initial value and ${\bar t}$ denotes the algorithm's iteration number.

Finally, we summarize the detailed steps to solve problem $\bf P5$ in Algorithm \ref{algorithmP10}.
Obviously, with the CSI decision made by the DROL, the SCA algorithm only requires a few iterations to solve the beamforming optimization problem and each iteration is with closed-form solutions \cite{guo2020weighted}. Consequently, the overall DROL method is an efficient algorithm only requiring polynomial complexity.
In comparison, the exhaustive search algorithm faces significant challenges. Firstly, it needs to generate $2^{K+Q}$ potential decisions, resulting in exponential complexity, particularly when dealing with numerous users/targets. Besides, it is impractical due to causality issues.

\begin{algorithm}[t]
\caption{FP/SCA Algorithm for problem \ref{eqproblem5}}\label{algorithmP10}
{{
\begin{algorithmic}[1]
 \STATE Initialize ${\bf{W}}^{\rm old}$, ${\bm{\mu }}^{\rm old}$, ${\bm{\alpha }}^{\rm old}$, ${\bm{\beta }}^{\rm old}$ and, ${\bm{\eta }}^{\rm old}$
\REPEAT 
\STATE Update $\bm \alpha$ by \eqref{eqwoptimize0};
 Update $\bm \beta$ by \eqref{eqwoptimize1};
 Update $\bm \mu$ by \eqref{eqwoptimize2};
 Update $\bf W$ by \eqref{eqwoptimize3replace};
 Update $\bm \eta$ by \eqref{eqwoptimize4}
\UNTIL{objective function converges
}
\end{algorithmic}}}
\end{algorithm}

\section{Simulation Results}

We consider that the system is operated on the carrier frequency of 5.89 GHz with a bandwidth of 10 MHz from IEEE 802.11p \cite{nguyen2017delay}. The numbers of transmit and receive antennas are $L_{\rm T}=64$ and $L_{\rm R}=32$.
The noise power spectral densities at the BS and communication users are both set as -174 dBm/Hz.
There are 6 communication users and the temporal correlations are set as $\rho_1=0.99$, $\rho_2=0.96$, $\rho_3=0.93$, $\rho_4=0.9$, $\rho_5=0.85$, and $\rho_6=0.8$. The path-loss is defined by $\bar \beta_k=74.2+16.11\log10(d_k^{\rm c}/d_o^{\rm c}),\forall k$, where  $d_o^{\rm c}=1$ m is the reference distance and $d_k^{\rm c}=4000$ m is the distance between the BS and user $k$ \cite{onubogu2014empirical}. The uplink transmit power for channel estimation  and the power of initial estimation error are set as   $P_k^u=30$ dBm and
$\varsigma _{{k}}^0=1/2\bar\beta_k$, respectively.
Besides, we consider 3 radar targets by setting the initial states as: ${\bf x}_1^0=[\pi/4 \;{\rm rad},150 \;{\rm m}, 30 \;{\rm m/s}]$, ${\bf x}_2^0=[\pi/3\;{\rm rad},150 \;{\rm m},30 \;{\rm m/s}]$, and ${\bf x}_3^0=[3\pi/4\;{\rm rad},150 \;{\rm m},30 \;{\rm m/s}]$. The state evolution noise covariances are set as $\epsilon_{q}^v=0.5\tilde\rho_qMT $, $\epsilon _{q}^d=0.5\tilde\rho_q\epsilon_{q}^v$, and $\epsilon_{q}^\theta=10^{-4}\tilde\rho_qMT$, respectively, where $\tilde\rho_0=0.05$, $\tilde\rho_1=1$, and $\tilde\rho_2=5$. As for the initial tracking error, we set
${\bf{M}}_q^0 = {\left( {\gamma _{q0}^{\rm{r}}} \right)^{ - 1}}{\rm{diag}}\left( {A_{q0}^\theta ,A_{q0}^d,A_{q0}^v} \right)$ with \eqref{escrlb} and ${{\gamma _{q0}^{\rm r}}}=\frac{P}{2(L_{\rm T}+L_{\rm R})(Q+N)}$.
Next, we set  ${\sigma _{{\rm{RCS,}}q}}=1$ and assume phase noise $\phi _q^n$ follows a uniform distribution between $\left[0,2\pi\right)$.
The  sub-gradient step-size parameter $\Delta_s$ is set as $0.005$. The remaining  parameters are given in Table \ref{table1} \cite{nguyen2017delay,mei2022multi}.

As for the DNN setup, we apply a fully connected DNN with the following layers: one input layer, four hidden layers with 1024, 1024, 258, and 64 hidden neurons, respectively, and one output layer. Since the ``LeakyReLU" function can address the issue of neuron death by allowing a small portion of negative inputs to pass through, leading to enhanced stability and generalization compared to the ``ReLU" function, we use ``LeakyReLU" with a leakage slope of 0.3 for hidden layers and ``Sigmoid" for the output layer \cite{xu2015empirical}.
The batch size, learning rate, and memory pool size are set as $\left| {{{\mathbb L}_n}} \right|=100$,  0.001, and 500, respectively. The updating intervals of ${\cal K}_{n }^{\rm{c}}$ and ${\cal K}_{n }^{\rm{r}}$ are given by  $\widetilde \Delta^{\rm c}=\widetilde \Delta^{\rm r}=4$, and the parameters in exploration probability are given by ${\cal A }^{\rm{c}}=1.9$ and ${\cal A }^{\rm{r}}=1$.

\begin{figure*}  \centering
  \begin{minipage}{.48\textwidth}
   \centering
   \includegraphics[width=\textwidth]{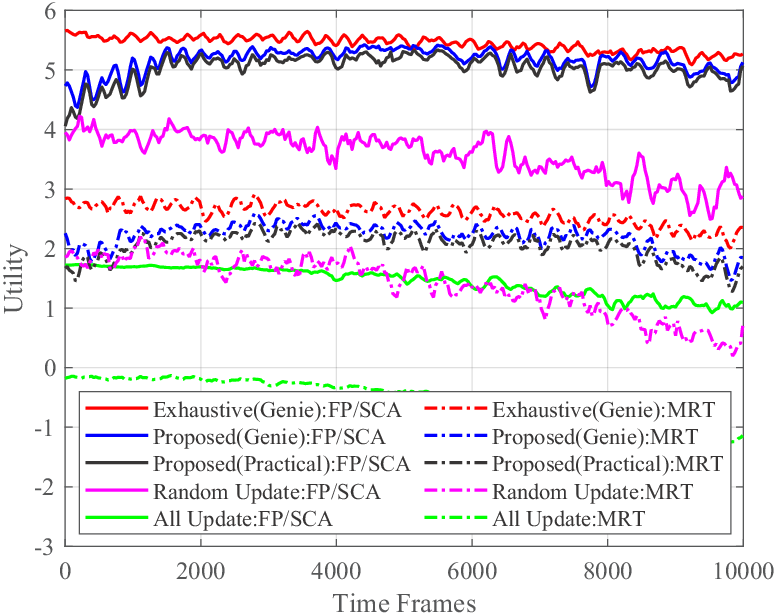}\vspace{-0.2cm}
   \caption*{  (a) System Utility }\label{figsimu11}
  \end{minipage}
  \begin{minipage}{.48\textwidth}
   \centering
   \includegraphics[width=\textwidth]{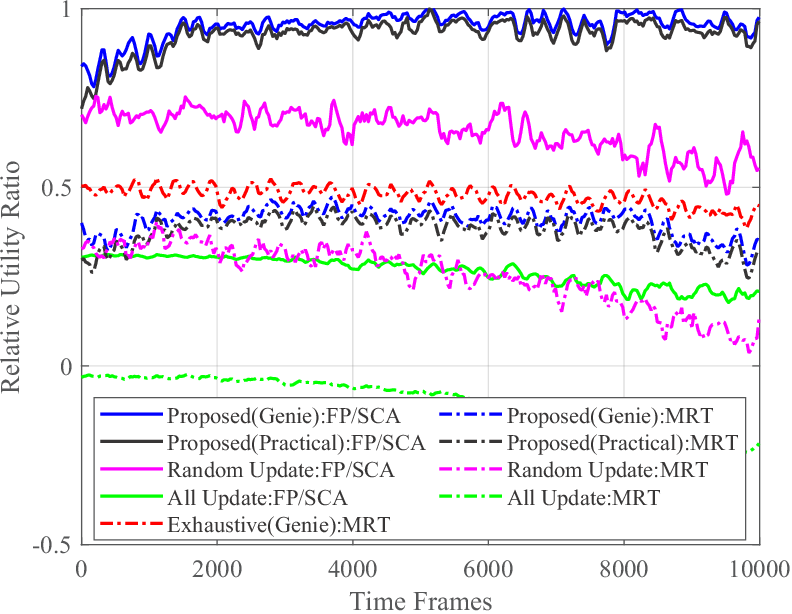}\vspace{-0.2cm}
    \caption*{  (b) Relative Utility Ratio}\label{figsimu12}
  \end{minipage}\\
\begin{minipage}{.48\textwidth}
   \centering
   \includegraphics[width=\textwidth]{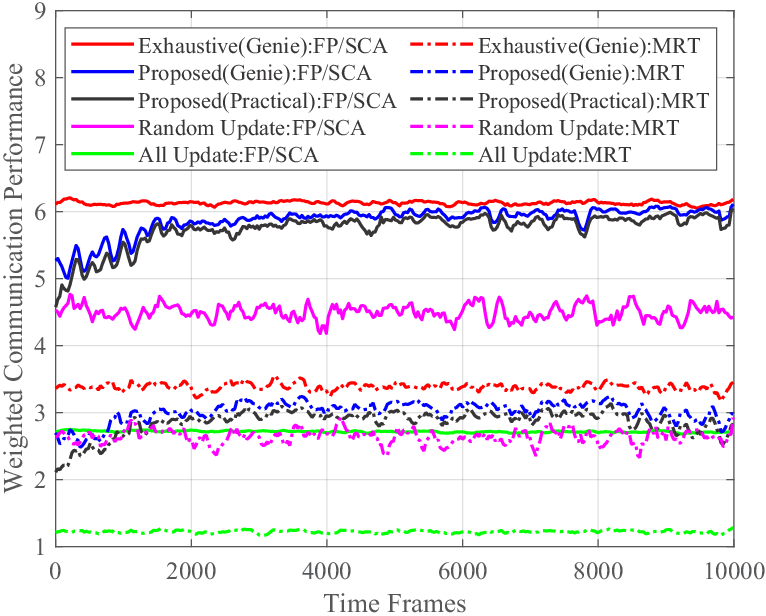}\vspace{-0.2cm}
    \caption*{  (c)  Weighted Communication Performance}\label{figsimu13}
  \end{minipage}
 \begin{minipage}{.48\textwidth}
   \centering
   \includegraphics[width=\textwidth]{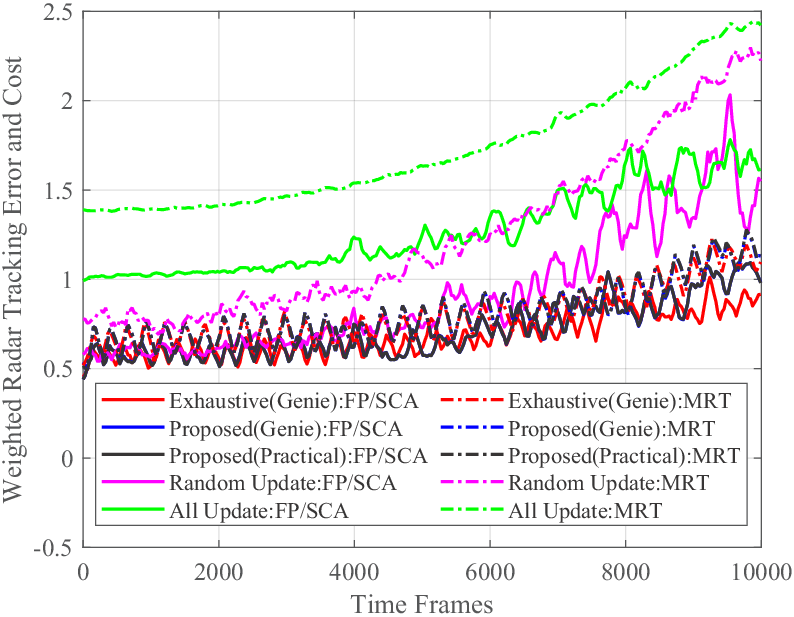}\vspace{-0.2cm}
    \caption*{  (d)  Weighted Radar Tracking Error and Cost}\label{figsimu14}
  \end{minipage}
  \caption{Performance comparisons among the proposed algorithm and other benchmarks with a moving average of 150 frames. }\label{figsimu1} \vspace{-0.3cm}
 \end{figure*}

In the following, we compare the system utility performance of our proposed scheme with that of the following schemes over 10,000 time frames. Note that in every 1000 frames, we set the channel state condition ${{\cal S}_n}$ in the other baseline schemes to the same value as that in our proposed scheme to ensure comparison fairness.
\begin{itemize}
\item Exhaustive (Genie):  In each frame, we perform an exhaustive search of all potential CSI updating decisions with FP/SCA-based or MRT beamforming  to obtain the system utility, i.e.,
      \begin{align}{\cal U}_n^{{\rm{Exhaustive}}} = \mathop {\max }\limits_{\scriptstyle{\bf{a}}_n^{\rm{c}} \in {\left\{ {0,1} \right\}^K}\hfill\atop
\scriptstyle{\bf{a}}_n^r \in {\left\{ {0,1} \right\}^Q}\hfill} {\cal U}\left( {{\bf{a}}_n^{\rm{c}},{\bf{a}}_n^{\rm{r}},{{\bf{W}}_n}} \right).
\end{align}
 Specifically, we only count the training overhead in the selected CSI updating decision and ignore that in the exploration of other decisions. This baseline uses minimal overhead to determine optimal decisions. Although it is impractical due to the causality issue, it can be applied to validate the performance of other algorithms.
  In addition, the MRT beamforming is given by ${\bf{w}}_k^n = \sqrt {\frac{P}{K}} \frac{{{\bf{\hat g}}_k^n}}{{\left\| {{\bf{\hat g}}_k^n} \right\|}}$ \cite{chen2023impact}.
\item Random Update: In each frame, we re-estimate the CSI of all users/targets randomly with a probability of 0.5, and then optimize beamforming matrices using FP/SCA or MRT methods.

\item All Update: In each frame, we re-estimate the CSI of all users/targets, and then optimize beamforming matrices using FP/SCA or MRT methods.
\end{itemize}

\begin{figure*}  \centering
  \begin{minipage}{.5\textwidth}
   \centering
   \includegraphics[ width=\textwidth, height= 6.2cm]{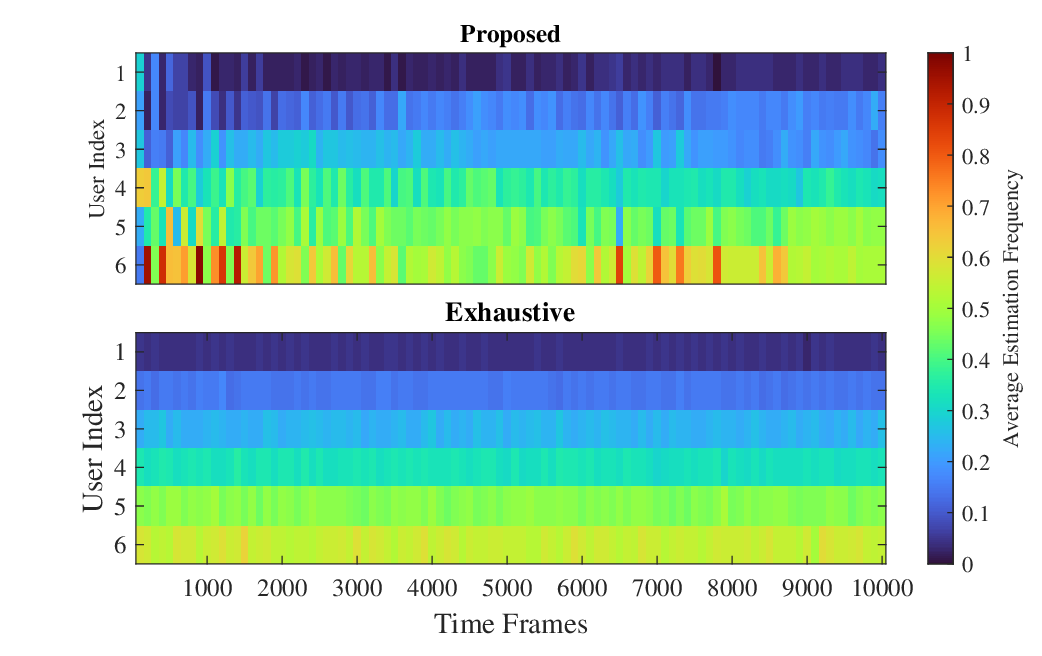}
   \caption*{  (a) Average communication CSI estimation frequency}\label{figsimu21}
  \end{minipage}
  \begin{minipage}{.48\textwidth}
   \centering
   \includegraphics[ width=\textwidth, height= 6.2cm]{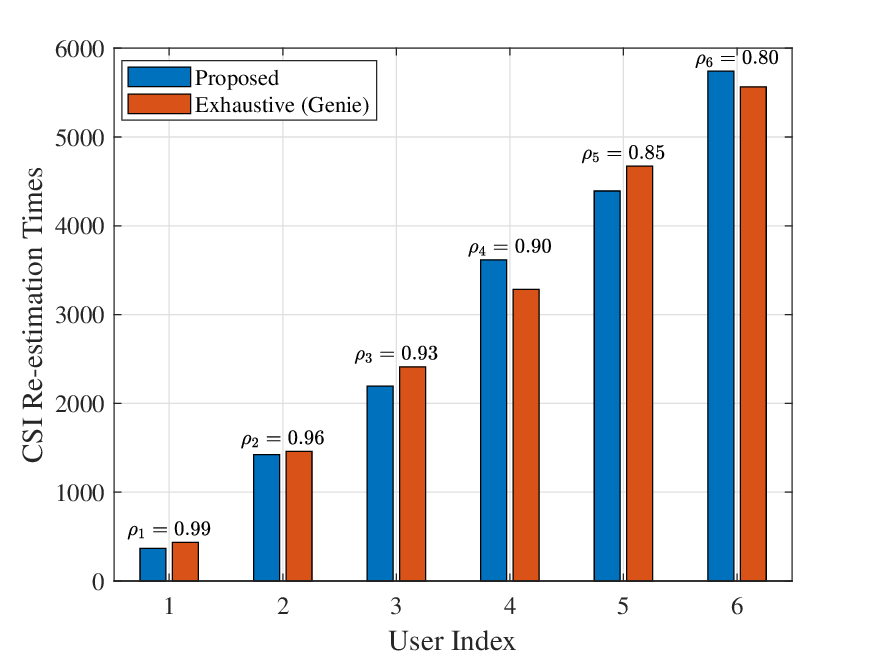}
    \caption*{  (b) Communication CSI re-estimation times in 10,000 frames}\label{figsimu22}
  \end{minipage}
  \caption{The impact of channel temporal coefficient on average communication CSI estimation frequency. Note that the channel temporal correlation coefficients are modeled in descending order according to the user index, i.e., ${\rho _k} > {\rho _{k'}}$ if $k < k'$.}\label{figsimu2}
 \end{figure*}

Fig. \ref{figsimu1} illustrates performance comparisons in each time frame: (a) system utility
${\cal U}\left( {{\bf{a}}_n^{\rm{c}},{\bf{a}}_n^{\rm{r}},{{\bf{W}}_n}} \right)$, (b) relative utility ratio
$\frac{{{\cal U}\left( {{\bf{a}}_n^{\rm{c}},{\bf{a}}_n^{\rm{r}},{{\bf{W}}_n}} \right)}}{{\cal U}_n^{\rm Exhaustive}}$,
 (c) weighted communication performance $\sum\nolimits_{k = 1}^K {w_k^{\rm{c}}C_k^n} $, and (d) weighted radar tracking error and cost  $\sum\nolimits_{q = 1}^Q {w_q^{\rm{r}}R_q^n} $.
\begin{itemize}
\item
From Fig. \ref{figsimu1}(a), we observe that each scheme using the FP/SCA  method outperforms the MRT method significantly.
This is because the FP/SCA method can approximate the non-convex problem as a convex problem using the first-order Taylor series expansion, and solve the approximated problem iteratively until it converges to the stationary solutions \cite{chen2019resource}.
Besides, it validates the effectiveness of the optimization method and the importance of beamforming design.
Then, we know the proposed algorithm increases initially and then decreases, in contrast to other schemes where the average utility consistently decreases.
This trend is attributed to the tracking error that increases over time frames as targets move away from the base station, resulting in system utility reduction.
However, the proposed scheme leverages the experiences to train DNN parameters to achieve better CSI updating decisions.
Thus, the utility of this scheme initially increases but decreases subsequently due to increased target distance.
Despite this, the proposed algorithm gradually approaches the exhaustive method and significantly outperforms other benchmarks.
Specifically,  Fig. \ref{figsimu1}(b) shows that the proposed algorithm can achieve more than 90\% performance of the exhaustive method when $n\ge2000$, which validates the effectiveness and convergence performances.

\item From Fig. \ref{figsimu1}(c), we know the average achievable communication rate of the genie-aided proposed scheme is better than that of the practical scheme, this is because the training overhead of exploring other CSI updating decisions is ignored in the former one.
Then, it shows that both of them gradually increase to approach the performance of exhaustive search, thanks to the better estimation decision made in the DNN training framework in each time frame.
Besides, due to the stationary communication channel, the rates of all algorithms remain nearly constant finally.
\item  From Fig. \ref{figsimu1}(d), we know that the weighted radar performances of all algorithms increase with the time frame. This is due to the increased tracking distance of the target. Besides, the proposed algorithm can achieve similar performance to the exhaustive search scheme and the growth rate is lower than other benchmarks, which further demonstrates its effectiveness.
    Finally, we know the random scheme gradually approaches the performance of all update schemes, which implies that the CSI updating frequencies should increase with the tracking distance.
\end{itemize}

\begin{figure*}\centering
  \begin{minipage}{.5\textwidth}
   \centering
   \includegraphics[width=\textwidth, height= 6.2cm]{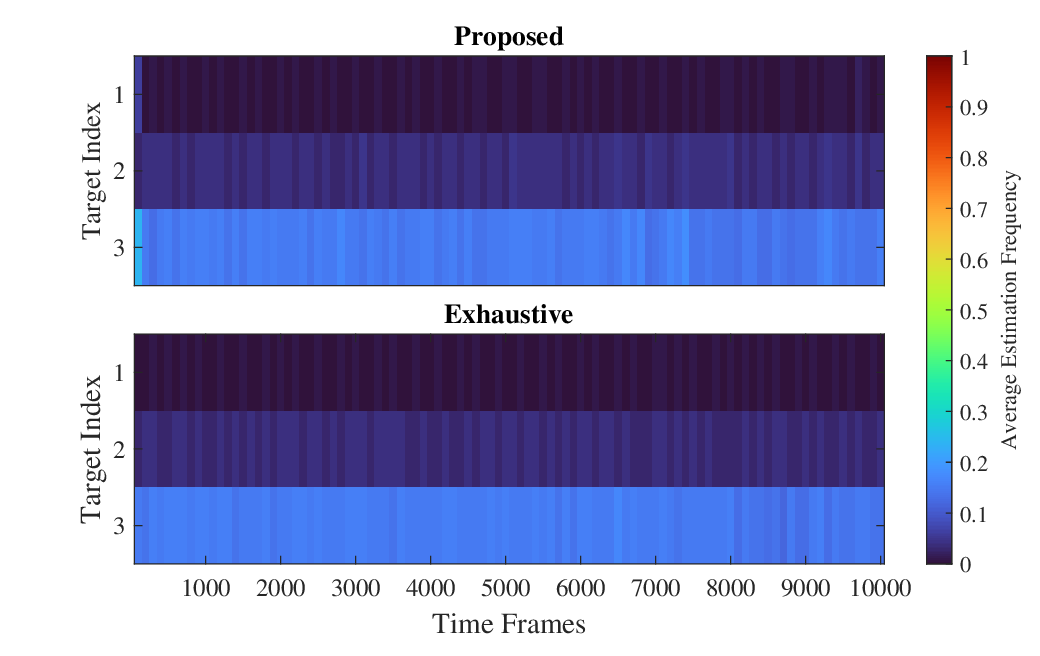}
   \caption*{ (a) Average radar CSI estimation frequency}\label{figsimu31}
  \end{minipage}
  \begin{minipage}{.48\textwidth}
   \centering
   \includegraphics[width=\textwidth, height= 6.2cm]{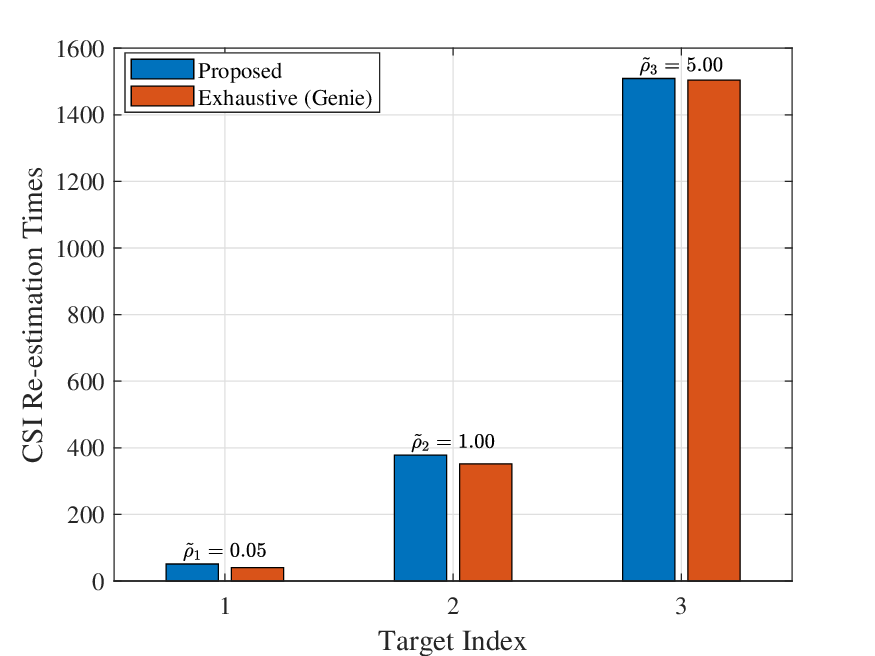}
    \caption*{ (b) Radar CSI re-estimation times in 10,000 frames  }\label{figsimu33}
  \end{minipage}
 \caption{The impact of state evolution noise covariance on average radar CSI estimation frequency. Note that the related noise power parameters are modeled in ascending order according to the target index, i.e., ${\tilde \rho _q} < {\tilde \rho _{q'}}$ if $q < q'$.}\label{figsimu3}
 \end{figure*}

\begin{figure}[t]
\centering
{\center\includegraphics[width=0.49\textwidth]{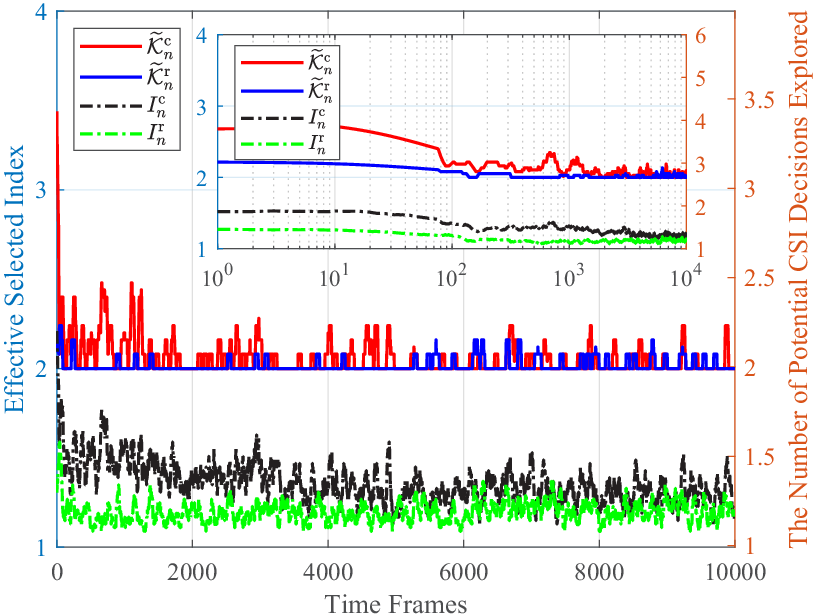} }
\caption{Average number of potential CSI decisions explored and the selected index of the optimal CSI decision in each frame over a moving average of 50 frames. } \label{figsim4}
\end{figure}

\begin{figure}[t]
\centering
{\center\includegraphics[width=0.49\textwidth]{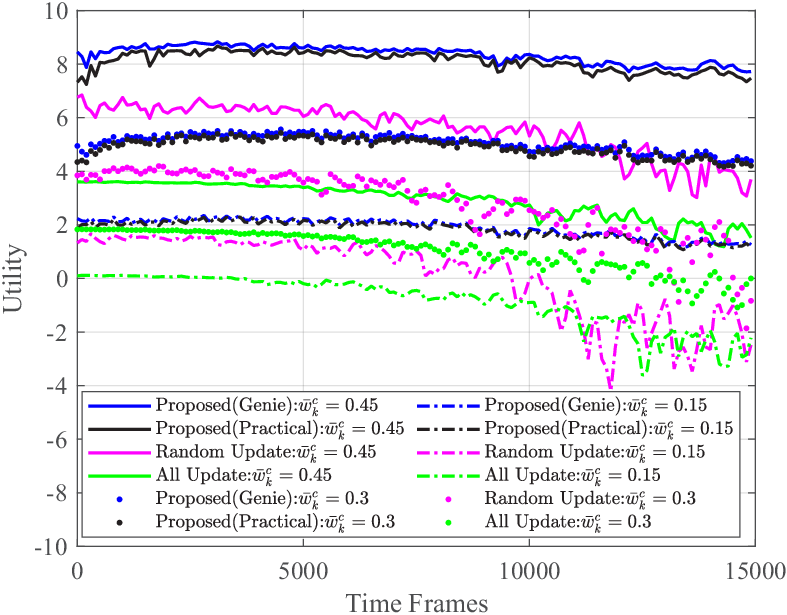} }
\caption{Comparison of system utility performance under varying weights averaged over a moving average of 150 frames.  } \label{figsimL}
\end{figure}

Fig. \ref{figsimu2} shows the impact of the channel temporal coefficient on communication CSI estimation frequency.
To clearly display the value of ${ {a}} _{n}^{\rm c}\left[ k \right]$ for each user in each frame,  we introduce the concept of average estimation frequency, which is defined as the average value of ${ {a}} _{n}^{\rm c}\left[ k \right]$ over every 50 time frames.
From Fig. \ref{figsimu2}(a), we observe that the CSI estimation frequency increases as the temporal correlation coefficient decreases. This is because a higher temporal correlation implies a lower variation between two successive frames, resulting in a more accurate CSI that can be achieved by prediction, thus the need for re-estimation can be reduced.
Besides, the communication CSI with a lower temporal coefficient requires a higher estimation frequency at the beginning.
Then, the CSI estimation frequency will decrease and remain almost constant. This is because the proposed algorithm needs to explore more frequently for the lower temporal correlated CSI until the DNN converges.
Once the DNN converges, the CSI will be approximately periodically re-estimated based on the learned individual stationary channel temporal correlation characteristics. This periodic estimation also appears in the exhaustive method. From Fig. \ref{figsimu2}(b),  the CSI re-estimation times of each user in the proposed algorithm are similar to that in the exhaustive method, and these numbers also increase as the correlation coefficient decreases.

Fig. \ref{figsimu3} shows the impact of channel state evolution noise covariance on radar CSI estimation frequency.
We observe that radar CSI estimation frequency and the radar CSI re-estimation times increase with the noise power.
The reason is that a higher noise power would increase the prediction error during the tracking process, resulting a higher estimation frequency.
Compared with Fig. \ref{figsimu2}, the periodic radar CSI estimation is more distinct than that in communication CSI estimation. It can be intuitively explained that the radar CSI only includes the mobility information, the evolution of which is more stationary than the whole channel response during communication CSI evolution.
 Finally, the total number of CSI estimations in the proposed algorithm is similar to that in the exhaustive method,  which can prove that the overhead has been successfully reduced. 

Fig. \ref{figsim4} shows the average number of potential CSI decisions explored and the selected effective index of the optimal CSI decision in each frame with a moving average of 50 frames, where the effective selected index is calculated by ${\rm{mod}}\left( {I_n^x,{\cal K}_n^x} \right)$ and $I_n^x$ denotes the index of the best solution ${\bf a}_{n,i}^{\rm x}$ of solving  $\bf P3$ in the $n$-th frame.
The smaller plot shows the same performance using a logarithmic scale. We observe that the average total number of potential CSI decreases with time, also the effective selected index decreases quickly and converges to near 1. This validates the effectiveness of the adaptive method and the convergence performance.

Fig. \ref{figsimL} illustrates the impact of varying performance weights on the system utility in each frame.
To further assess algorithm robustness, we consider different distances for each user, denoted as $d_k^{\rm c}=32000+200(k-1)$ meters, and different distance and velocity values for the targets, specified as
 ${\bf x}_1^0=[\pi/4 \;{\rm rad},110 \;{\rm m}, 20 \;{\rm m/s}]$,
 ${\bf x}_2^0=[\pi/3\;{\rm rad},140 \;{\rm m},30 \;{\rm m/s}]$, and
 ${\bf x}_3^0=[3\pi/4\;{\rm rad},170 \;{\rm m},40 \;{\rm m/s}]$.
From this figure,  we observe that the proposed algorithm outperforms other baseline methods, which validates its robustness and effectiveness.

\section{Conclusions}
This paper focuses on solving the challenging problem of ISAC systems: when the communication/radar CSI of each user/target should be re-estimated according to individual channel time-varying characteristics and system performance demands.
 Hence, we propose an intermittent communication and radar CSI estimation scheme with adaptive individual intervals for each user/target in an ISAC MISO system.
Specifically, this scheme jointly optimizes binary CSI re-estimation or prediction decisions and beamforming matrices to reduce training costs and improve system utility.
To further solve the causality and complexity issues,  the DROL framework is proposed to implement an online DNN to learn binary CSI updating decisions from the experiences.
Then, with the learned updating decisions, the FP/SCA algorithm is proposed to solve the remaining beamforming problem efficiently.
Finally, the simulation results validate the effectiveness of the proposed algorithm and show that the communication CSI with higher temporal correlation and the radar CSI with lower state evolution noise covariance require lower CSI estimation frequencies.
This outcome can be used to determine channel update frequencies in practical ISAC applications, reducing estimation overhead without requiring complex optimization procedures.
Also, note that the proposed algorithms are applicable for rectangular pulse shapes and can also be adapted for other pulse shapes.

\appendix
\subsection{Proof of Theorem \ref{theorem1}} \label{theorem1proof}
In this part, we perform the posterior estimation of radar CSI and derive its  PCRB performance.

Upon denoting ${\bf{\tilde y}}_n^{\rm{r}}\left[ {mb} \right] = {\bf{y}}_n^{\rm{r}}\left( t \right)\left| {_{t = (M_{1}^n+m)T + {T_{cp}} + \frac{{{b}}}{{B}}T_o}} \right.$ in \eqref{eq13},  we have
 \begin{align}
{\bf{\bar y}}_n^{\rm{r}}\left[ {mb} \right]& = \frac{1}{B}\sum\nolimits_{\tilde b = 0}^{B - 1} {{\bf{\tilde y}}_n^{\rm{r}}\left[ {m\tilde b} \right]{e^{ - {\rm{j}}2\pi \tilde b\frac{b}{B}}}} \nonumber \\
 &= \sqrt {{L_{\rm{R}}}{L_{\rm{T}}}} \sum\nolimits_{q = 1}^Q {\left[ {\bar \alpha _q^n{e^{{\rm{j}}\phi _q^n}}{{\bf{v}}_{\rm{R}}}\left( {\theta _q^n} \right){e^{{\rm{j}}2\pi \nu _q^nmT}}} \right.}\nonumber  \\
&\qquad\qquad\times\left. {\chi _q^n\left[ {mb} \right]{e^{ - {\rm{j2}}\pi b{\Delta _f}\tau _q^n}}} \right] + {\bf{\bar u}}_n^{\rm{r}}\left[ {mb} \right], \label{eqmulticase}
\end{align}
where ${\bf{\bar u}}_n^{\rm{r}}\left[ {mb} \right] \sim{\cal  CN}\left( {0,{\sigma } {{\bf{I}}_{{L_R}}}} \right)$ is the noise and $\sigma ={\Delta_f \tilde \delta}$. Here,  $\chi _q^n\left[ {mb} \right] = \sum\nolimits_{k = 1}^K  {\bf{v}}_{\rm{T}}^H\left( {\theta _q^n} \right){\bf{w}}_k^n\tilde s_k^n\left[ {mb} \right]$.

Then, by applying Periodogram-based estimation algorithm \cite{braun2014ofdm},  ${\bf{x}}_q^n $ can be estimated as
\begin{align}\bar{\bf{x}}_q^n
 = \mathop {\arg \max }\limits_{\left( {\theta ,\nu ,\tau } \right)} \left| {\sum\nolimits_{m = 0}^{M_2^n - 1} {\sum\nolimits_{b = 0}^{B - 1} {{\bf{v}}_{\rm{R}}^H\left( \theta  \right){\bf{\bar y}}_n^{\rm{r}}\left[ {mb} \right]} } } \right.\nonumber\\
{\left. {{{\left( {\chi _q^n\left[ {mb} \right]} \right)}^*}{e^{j2\pi \left( {b{\Delta _f}\tau  - mT\nu } \right)}}} \right|^2} \label{eq1s}.
\end{align}

When ${L_{\rm{R}}} \to \infty$,  the inner product of two receive antenna steering vectors with different angles is approximately zero, i.e., ${\bf{v}}_{\rm{R}}^H\left( {{\theta_q}} \right){{\bf{v}}_{\rm{R}}}\left( {{\theta _{q'}}} \right) = 0$ if $1 \leq q \neq q' \leq Q$. We also assume that the targets are widely separated in the surveillance region \cite{yan2015simultaneous}.
Based on these assumptions, it has been shown in \cite{garcia2014resource1} that the CRB matrix for multiple targets is a block diagonal matrix, and the CRB for each target  can be approximately and individually derived from the single target case of \eqref{eqmulticase}, i.e.,
\begin{align}
{\bf{\bar y}}_{qn}^{\rm{r}}\left[ {mb} \right] = \sqrt {{L_{\rm{R}}}{L_{\rm{T}}}} \bar \alpha _q^n{e^{{\rm{j}}\phi _q^n}}{{\bf{v}}_{\rm{R}}}\left( {\theta _q^n} \right){e^{{\rm{j}}2\pi \nu _q^nmT}}\nonumber\\
 \times \chi _q^n\left[ {mb} \right]{e^{{\rm{ - j2}}\pi b{\Delta _f}\tau _q^n}} + {\bf{\bar u}}_{qn}^{\rm{r}}\left[ {mb} \right] \label{singlecase},
\end{align}
where  ${\bf{\bar u}}_{qn}^{\rm{r}}\left[ {mb} \right] \sim{\cal  CN}\left( {0,{\sigma }{{\bf{I}}_{{L_R}}}} \right)$ is the  noise term.

From \eqref{singlecase}, the CRBs on the estimation MSEs of $\theta _q^n$, $d _q^n$, and $v _q^n$  can be approximated by \cite{chen2023impact}
\begin{subequations}\label{escrlb}
\begin{align}
\!\!\!\sigma _{\theta _q^n}^2& = \frac{{6\sigma }}{{\Xi_q^n {{\cos }^2}\left( {\theta _q^n} \right)\left( {L_{\rm{R}}^2 - 1} \right)\gamma _{qn}^{\rm{r}}}} \buildrel \Delta \over = \frac{{A_{qn}^\theta }}{{\gamma _{qn}^{\rm{r}}}},\\
\!\!\!\sigma _{d_q^n}^2& = \frac{{3c_0^2\sigma }}{{8\Xi_q^n \Delta _f^2\left( {{B^2} - 1} \right)\gamma _{qn}^{\rm{r}}}} \buildrel \Delta \over = \frac{{A_{qn}^d}}{{\gamma _{qn}^{\rm{r}}}},\\
\!\!\!\sigma _{v_q^n}^2 &= \frac{{3c_0^2\sigma }}{{8\Xi_q^n {{\left| {T{f_c}\cos \left( {\bar \theta _q^n} \right)} \right|}^2}\left( {{{(M_2^n)}^2} - 1} \right)\gamma _{qn}^{\rm{r}}}} \buildrel \Delta \over = \frac{{A_{qn}^v}}{{\gamma _{qn}^{\rm{r}}}},
\end{align}
\end{subequations}
respectively, where $\gamma _{qn}^{\rm{r}} = {\mathbb{E}}({\left| {\chi _q^n\left[ {mb} \right]} \right|^2}) =  {\sum\nolimits_{k = 1}^K {{{\left| {{\bf{v}}_{\rm{T}}^H\left( {\theta _q^n} \right){\bf{w}}_k^n} \right|}^2}} } $ and $\Xi_q^n  = {\left| {\bar \alpha _q^n\pi } \right|^2}BM_2^n{L_{\rm{R}}}{L_{\rm{T}}}$.

Note that the above derived CRB provides a lower bound for the variances of any unbiased estimators \cite{chen2023impact}. However, this bound only considers the current received echo signals and does not consider temporal correlations of radar CSI, i.e., \eqref{eqtcorrR}. In tracking scenarios, it is necessary to derive the PCRB, which considers both the measurements and temporal correlations.
To do so, we combine \eqref{eqtcorrR}, \eqref{eq1s}, and \eqref{escrlb} to get the  posterior estimation of radar CSI, i.e.,
 \begin{align}
\left\{ {\begin{array}{*{20}{l}}
{\bf{x}}_q^n = \Gamma\left( {{\bf{x}}_q^{n - 1}} \right) + {\bm \epsilon}_q^{n - 1},\\
\bar{\bf{x}}_q^n = {\bf{x}}_q^n + {\bm \delta} _q^n,
\end{array}} \right.\label{eqstate}
\end{align}
where
 ${\bm \delta} _q^n\sim {\cal CN}\left( {0,\frac{{\bf{\Sigma }}_{qn}^\delta}{{\gamma _{qn}^{\rm r}}}} \right)$ with  matrix
 ${\bf{\Sigma }}_{qn}^\delta  = {\rm{diag}}\left( {{{A_{qn}^\theta }},{{A_{qn}^d}},{{A_{qn}^v}} } \right)$.
By applying EKF algorithm in \cite{ristic2003beyond}  with \eqref{eqrp1}, the predict and posterior estimations of ${\bf{\hat x}}_q^n$ are given by \eqref{eq19}.
From \cite{liu2020radar}, the PCRBs are given in \eqref{eq20}.

 \subsection{ Order-Preserving Quantization Method}\label{orderproof}
Here, we introduce the order-preserving quantization   to  generate $\widetilde {\cal K}_n^{\rm x}$ integer solutions from ${\widetilde {\bf a}_n^{\rm{x}}}$.
Specifically, denoting the $i$-th integer solution of ${\widetilde {\bf a}_n^{\rm{x}}}$ by ${ {\bf a}_{n,i}^{\rm{x}}}$ and its $l$-th entry by ${ { a}_{n,i}^{\rm{x}}}\left[l\right]$, the condition that $ { a}_{n,i}^{\rm x} \left[ \ell \right]\ge { a}_{n,i}^{\rm x} \left[ \ell' \right]$ if  $\widetilde { a}_{n}^{\rm x} \left[ \ell \right]\ge \widetilde { a}_{n}^{\rm x}\left[ \ell' \right]$ for all $\ell,\ell'\in\left\{1,\cdots,X\right\}$ should be guaranteed to preserve the ordering during quantization \cite{huang2019deep}.
Therefore,  the first feasible action is given by
\begin{align}
{ a}_{n,1}^{\rm x}\left[ \ell \right] = \left\{ {\begin{array}{*{20}{l}}
{1,\;{\rm{if}}\;\widetilde{ a}_{n}^{\rm x}\left[ \ell \right] \ge 0.5},\\
{0,{\rm{otherwise}}},
\end{array}} \right.\label{eqfirst1}
\end{align}
and the next $(\widetilde {\cal K}_n^{\rm x}-1)$ integer solutions ${\bf a}_{n,i}^{\rm x}$ for $2\le i \le \widetilde {\cal K}_n^{\rm x}$ can be given by
\begin{align}
a_{n,i}^{\rm{x}}\left[ \ell  \right] = \left\{ \begin{array}{l}
1,\;{\rm{if}}\;
{{{\widetilde a}}_n^{\rm{x}}\left[ \ell  \right] > {{\widetilde a}}_n^{\rm{x}}\left[ {\left( {i - 1} \right)} \right],} {\rm{or}}\\
{\;\left\{
{{{\widetilde a}}_n^{\rm{x}}\left[ \ell  \right] = {{\tilde a}}_n^{\rm{x}}\left[ {\left( {i - 1} \right)} \right]\;{\rm{and}}\;{{\widetilde a}}_n^{\rm{x}}\left[ {\left( {i - 1} \right)} \right] \le 0.5\;},\right\},} \\
0,{\rm{if}}\;
{{{\widetilde a}}_n^{\rm{x}}\left[ \ell  \right] < {{\widetilde a}}_n^{\rm{x}}\left[ {\left( {i - 1} \right)} \right],}{\rm{or}}\\
{\;
\left\{ {{{\widetilde a}}_n^{\rm{x}}\left[ \ell  \right] = {{\widetilde a}}_n^{\rm{x}}\left[ {\left( {i - 1} \right)} \right]\;{\rm{and}}\;{{\widetilde a}}_n^{\rm{x}}\left[ {\left( {i - 1} \right)} \right] > 0.5\;} \right\},}
\end{array} \right.
\label{eqfirst2}
\end{align}
where $ {{{\widetilde {{a}}}_n^{\rm{x}}}\left[ {\left(i \right)} \right]}$ is the $i$-th ordered element of $\widetilde {\bf a}_n^{\rm{x}}$ compared with the distance to 0.5, i.e., $\left| {{{\widetilde {{a}}}_n^{\rm{x}}}\left[ {\left( i \right)} \right] - 0.5 } \right| \le \left| {{{\widetilde {{a}}}_n^{\rm{x}}}\left[ {\left( j \right)} \right] - 0.5 } \right|$ if $1 \le i \le j \le X$.

\subsection{Prox-linear Method to Update $\bf W$} \label{appendixW}
From \cite{guo2020weighted}, the approximated problem of optimizing  $\bf W$ is
 \begin{align}
\!  \!\min\limits_{\bf W} \sum\limits_{k = 1}^K {{\mathop{\rm Re}\nolimits} \left( {{\bf{f}}_k^H\left( {{{\bf{w}}_k} - {{{\bf{\tilde w}}}_k}} \right)} \right)}  + \frac{\tilde f}{2}{\left\| {{{\bf{w}}_k} - {{{\bf{\tilde w}}}_k}} \right\|^2}\;
{\rm{s}}.{\rm{t}}.\eqref{eqP1sj1},\tag{\bf P7}
  \end{align}
 where ${{{\bf{\tilde w}}}_k} = {{\bf{w}}_k^{\rm old}} + \dot \varsigma  \left( {{{\bf{w}}_k^{\rm old}} - {\bar {\bf{w}}_k}^{\rm old}} \right)$ is an extrapolated point,   $\dot \varsigma  \ge0$  is the extrapolation weight, and ${\bar {\bf{w}}_k}^{\rm old}$ is the value of  ${ {\bf{w}}_k}$ before it was updated to ${{\bf{w}}_k^{\rm old}}$ in the last second iteration. Besides, $\tilde f>0$ and the gradient ${\bf{f}}_k $ is given by
 \begin{align}
 {{\bf{f}}_k} &=- \frac{{\cal F}\left(   {\bm{\alpha }},{\bm{\beta }},{\bm{\mu }},{\bf{W}},{\bm \eta} \right)}{{\partial {{\bf{w}}_k}}}\left| {_{{{\bf{w}}_k} = {{{\bf{\tilde w}}}_k}}} \right.\nonumber\\
 & = 2\left(- {{{\bf h}_{k}^n} + \sum\nolimits_{j = 1}^K {{{\left| {\beta _j^{{\rm{new}}}} \right|}^2}{{{\bf{\hat g}}}_j^n}({\bf{\hat g}}_j^n)^H{{{\bf{\tilde w}}}_k}} } \right).
  \end{align}
Then,   ${\bf P7}$ is convex with respect to $\bf W$, thus its optimal solution can be obtained by using the Lagrangian method, i.e.,
  \begin{align}
{{\bf{w}}_k^{\rm new}} &= \frac{1}{{\tilde f - 2{\tilde \lambda}}}\left( {{\tilde f}{{{\bf{\tilde w}}}_k}  - {{\bf{f}}_k}} \right), \label{eqwoptimize3replace}\\
\tilde \lambda  &= \frac{{\tilde f}}{2} - \frac{1}{2}\sqrt {\frac{1}{P}\sum\nolimits_{k = 1}^K {{{\left\| {\tilde f{\bf{\tilde w}}_k- {{\bf{f}}_k}} \right\|}^2}} }. \end{align}
There is no matrix inverse operation and additional iteration of linear search, thus the complexity can be reduced.
As for  parameters $\tilde f$ and $\dot \alpha $, we apply the Lipschitz constant of $ {{\bf{f}}_k}$ to set $\tilde f$, i.e.,  $\tilde f = 2\left\| {\sum\nolimits_{j = 1}^K {{{\left| {{\beta _j^{\rm new}}} \right|}^2}{{{\bf{\hat g}}}_j^n}({\bf{\hat g}}_j^n)^H} } \right\|$. Then, the extrapolation weight can be set as $\dot \varsigma  = \min \left( {\frac{{\dot d - 1}}{{\ddot d}},0.9999\sqrt {\frac{{\ddot f}}{{\tilde f}}} } \right)$, $\dot d = \frac{1}{2}\left( {1 + \sqrt {1 + 4{{\ddot d}^2}} } \right)$, where ${\ddot d}$ and  ${\ddot f}$ are the values of ${\dot d}$ and  ${\dot f}$  adopted in previous iteration.

\ifCLASSOPTIONcaptionsoff

\fi
  \bibliography{SPT}
\bibliographystyle{IEEEtran}

\begin{IEEEbiography}[{\includegraphics[width=1in,height=1.25in,clip,keepaspectratio]{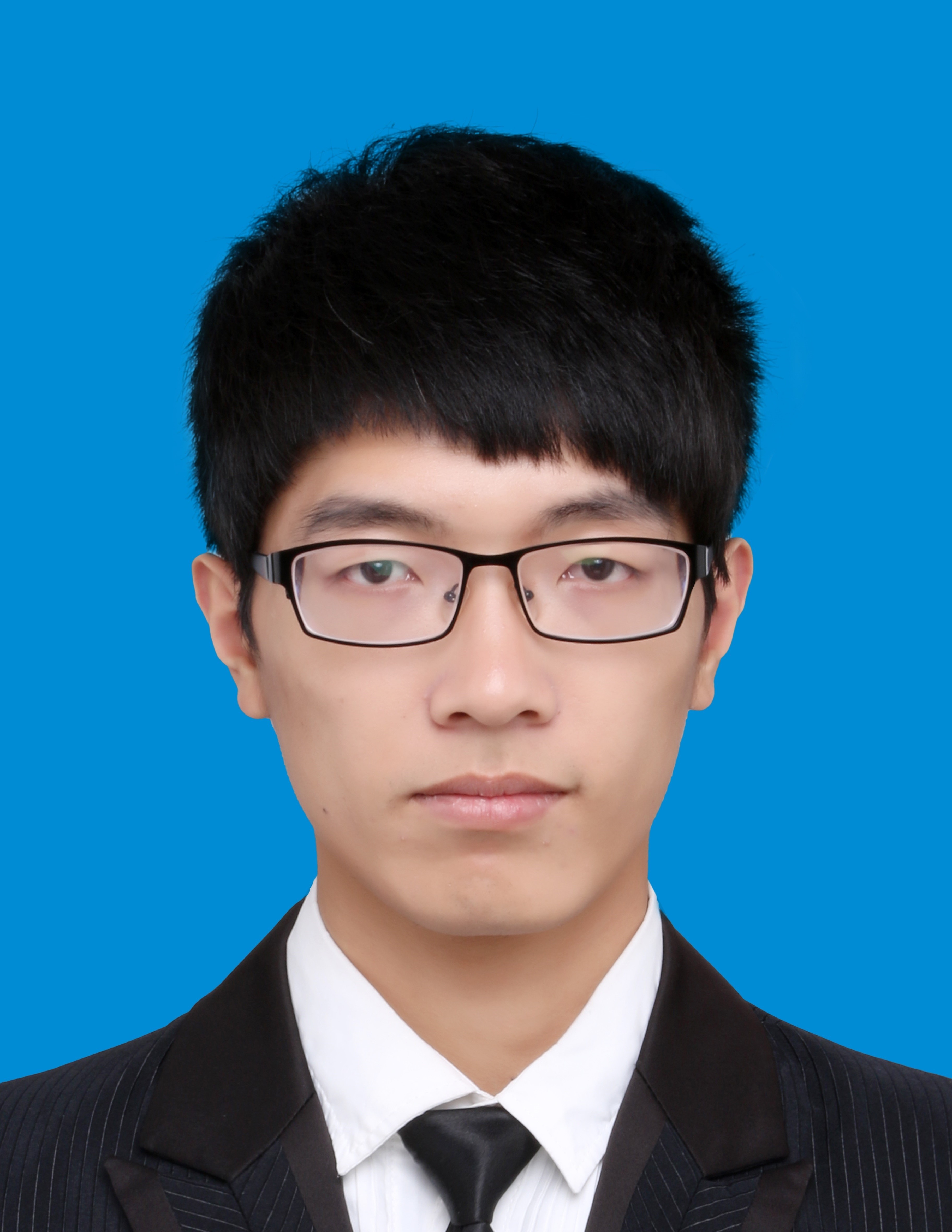}}]{Jie Chen} received the B.S. degree in communication engineering from Chongqing University of Posts and Telecommunications, China, in 2016, and the Ph.D. degree from University of Electronic Science and Technology of China (UESTC), China, in 2021.
He was a Visiting Student Research Collaborator at the University of Toronto, Canada, from 2019-2020.
He is currently a Post-Doctoral Research Fellow at the Department of Electrical and Computer Engineering, Western University, London, ON, Canada.
His research interests include integrated sensing and communications, transceiver design for Internet-of-Things, and machine learning for wireless communications. He received the  Journal of Communications and Information Networks (JCIN) Best Paper Award in 2021.
\end{IEEEbiography}

\begin{IEEEbiography}[{\includegraphics[width=1in,height=1.25in,clip,keepaspectratio]{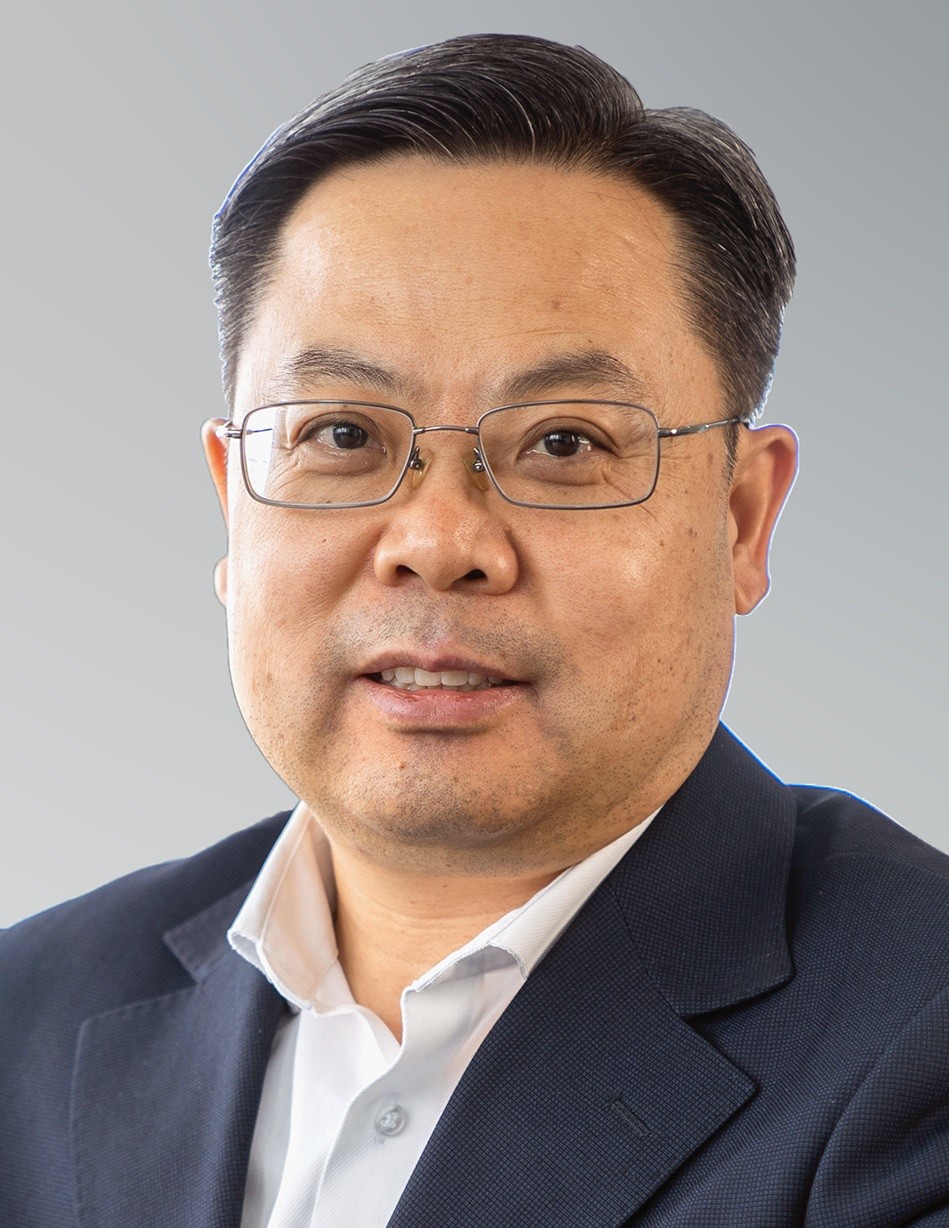}}]{Xianbin Wang} (Fellow, IEEE)  received his Ph.D. degree in electrical and computer engineering from the National University of Singapore in 2001.

He is a Professor and a Tier-1 Canada Research Chair in 5G and Wireless IoT Communications with Western University, Canada. Prior to joining Western University, he was with the Communications Research Centre Canada as a Research Scientist/Senior Research Scientist from 2002 to 2007. From 2001 to 2002, he was a System Designer at STMicroelectronics. His current research interests include 5G/6G technologies, Internet of Things, machine learning, communications security, and intelligent communications. He has over 600 highly cited journals and conference papers, in addition to over 30 granted and pending patents and several standard contributions.

Dr. Wang is a Fellow of the Canadian Academy of Engineering and a Fellow of the Engineering Institute of Canada. He has received many prestigious awards and recognitions, including the IEEE Canada R. A. Fessenden Award, Canada Research Chair, Engineering Research Excellence Award at Western University, Canadian Federal Government Public Service Award, Ontario Early Researcher Award, and nine Best Paper Awards. He is currently a member of the Senate, Senate Committee on Academic Policy and Senate Committee on University Planning at Western. He also serves on NSERC Discovery Grant Review Panel for Computer Science. He has been involved in many flagship conferences, including GLOBECOM, ICC, VTC, PIMRC, WCNC, CCECE, and ICNC, in different roles, such as General Chair, TPC Chair, Symposium Chair, Tutorial Instructor, Track Chair, Session Chair, and Keynote Speaker. He serves/has served as the Editor-in-Chief, Associate Editor-in-Chief, and editor/associate editor for over ten journals. He was the Chair of the IEEE ComSoc Signal Processing and Computing for Communications (SPCC) Technical Committee and is currently serving as the Central Area Chair of IEEE Canada.

\end{IEEEbiography}

\end{document}